\documentclass[twocolumn]{autart}
\usepackage{alltt}
\usepackage{color}
\usepackage{amsmath,amssymb}
\usepackage[square, comma, sort&compress, numbers]{natbib}
\usepackage{graphicx}
\usepackage{subfigure}
\usepackage{framed}

\definecolor{green2}{rgb}{0,0.62,0.17}
\definecolor{orange}{rgb}{1,0.5,0.0}
\definecolor{b}{rgb}{0,0,1}
\definecolor{r}{rgb}{1,0,0}

 \newcommand{\R}{\mathbb{R}}
 \newcommand{\N}{\mathbb{N}}

  \newcommand{\dn}{\mathbf{d}}
 \newcommand{\sign}{\mathrm{sign}}
 \newcommand{\diag}{\mathrm{diag}}

  \newcommand{\zero}{\mathbf{0}}

  \newcommand{\rank}{\mathrm{rank}}

\newtheorem{definition}{Definition}
\newtheorem{lemma}{Lemma}
\newtheorem{theorem}{Theorem}
\newtheorem{corollary}{Corollary}

\begin{document}
\begin{frontmatter}
\title{Filtering Homogeneous Observer for MIMO System}

\thanks[fi]{This work is supported by RSF under grant 21-71-10032.}
\author [1]{Xubin Ping},
\author[2]{Konstantin Zimenko},
\author [3]{Andrey Polyakov},
\author [3]{Denis Efimov}

\address[1]{Xidian University, Xi'an, 710071, China}
\address[2]{ITMO University, Saint-Petersburg, 197101, Russia}
\address[3]{Univ. Lille, Inria, CNRS, UMR 9189  CRIStAL, Centralle Lille, F-59000 Lille, France}

\begin{abstract}
Homogeneous observer for linear multi-input multi-output (MIMO) system is designed. A prefilter of the output is utilized in order to improve robustness of the observer with respect to measurement noises. The use of such a prefilter also simplifies tuning, since the observer gains in this case are  parameterized by a linear matrix inequality (LMI) being always feasible for observable system. In particular case, the observer is shown to be applicable in the presence of the state and the output bounded perturbations.  Theoretical results are supported by numerical simulations.
\end{abstract}
\end{frontmatter}
\section{Introduction}
Observation problem consists in a  reconstruction of the  state of a dynamical system  based on some output measurements. The system must satisfy certain conditions to be observability  \cite{HermannKrener1977:TAC}, \cite{Wonham1985:Book}, \cite{Isidori1995:Book}. An observer usually is a dynamical model, which state (or a part of the state vector) converges to the state of the observable system asymptotically or in a finite/fixed time. Various asymptotic \cite{Luenberger1964:TMI}, \cite{KhalilPraly2013:IJRNC} and  non-asymptotic \cite{KreisselmeierEngel2003:TAC}, \cite{Levant2003:IJC}, \cite{Cruz_etal2011:TAC}  observers are developed having  their own advantages and disadvantages.

Homogeneity is a dilation symmetry \cite{Zubov1958:IVM}, \cite{Kawski1991:ACDS}, \cite{BhatBernstein2005:MCSS}, \cite{Andrieu_etal2008:SIAM_JCO}, \cite{Polyakov2020:Book}.
If $\dn(s):\R^n\mapsto \R^n, s\in \R$  is a one-parameter group of dilations \cite{Kawski1991:ACDS}, \cite{Husch1970:Math_Ann}, \cite{FischerRuzhansky2016:Book}  in $\R^n$ and $g :\R^n\mapsto \R^n$ is a vector field on $\R^n$, then  the
 $\dn$-homogeneity of $g$ is its symmetry $g(\dn(s)\varepsilon)=e^{\nu s} \dn(s)g(\varepsilon), \forall s\in \R, \forall \epsilon\in \R^n$ with respect to the dilation $\dn$, where $\nu\in \R$ is the so-called homogeneity degree.
By \cite{Zubov1964:Book}, \cite{BhatBernstein2005:MCSS}, the homogeneity degree specifies a convergence rate of the asymptotically stable $\dn$-homogeneous  system $\dot \varepsilon=g(\varepsilon)$, e.g., the negative degree corresponds to finite-time convergence.
 Homogeneous systems are also known to be robust with respect to a rather large class of perturbations \cite{Ryan1995:SCL}, \cite{Hong2001:Aut}, \cite{Andrieu_etal2008:SIAM_JCO}.

 The so-called weighted homogeneous observers with $\dn(s)=\diag(e^{r_1s},...,e^{r_ns}),r_i>0$ are developed, e.g., in \cite{Levant2003:IJC}, \cite{BhatBernstein2005:MCSS}, \cite{Perruquetti_etal2008:TAC}, \cite{Andrieu_etal2008:SIAM_JCO}, \cite{CruzMoreno2017:Aut} for systems topologically equivalent to the chain of integrators. Extensions of the same ideas to the multi-output case can be found, for example, in  \cite{Lopez-Ramirez_etal2018:Aut}, \cite{Moreno2023:RNC}. Homogeneous finite-time observers
 are rather popular state estimation algorithms due to simplicity of digital implementation, better robustness and convergence properties. However,  the key  difficulty of their application is the absence of characterization of admissible gains of homogeneous observers despite  several attempts to establish it. Such a characterization is necessary for development of various tuning algorithms (e.g., based on attractive ellipsoids method \cite{PoyakTopunov2008:ARC}).
 It is known  \cite{BhatBernstein2005:MCSS}, \cite{Andrieu_etal2008:SIAM_JCO}, \cite{Perruquetti_etal2008:TAC} that any linear observer can be transformed (``upgraded'') to a homogeneous one provided that  the so-called  homogeneity degree is close to zero. In this particular case, the homogeneous observer has to be close to linear in order to share the same gains. However, most advanced homogeneous observers, such as  Levant's diffirentiator \cite{Levant1998:Aut}, correspond to essentially nonlinear cases with non-small homogeneity degrees. Existing schemes for parameters tuning of the high order Levant's differentiator  are  non-constructive \cite{Levant2003:IJC} (i.e., the set of admissible gains is unknown) and/or essentially nonlinear \cite{Moreno2022:TAC}.

The so-called filtering homogeneous observer is designed recently \cite{Jbara_etal2021:RNC} for the integrator chain. It combines the conventional homogeneous observer with a  homogeneous filter. Such a combination helps to decrease a sensitivity of the observer with respect to measurement noises. A procedure for a recursive design of possible observer's parameters is suggested. The procedure does not provide constructive restrictions to the observer parameters claiming that they have to be selected sufficiently large. A constructive (LMI-based) tuning of the particular filtering observer for the integrator chain is proposed recently in \cite{Nekhoroshikh_etal2022:CDC}. This paper generalizes the latter paper and designs a filtering homogeneous observer for linear MIMO systems.\vspace{-1mm}

\textit{Contributions}:\vspace{-3mm}
\begin{itemize}
 \item The gains of the observer are characterized by a Linear Matrix Inequality (LMI) being similar to  the LMI for the linear  observer design \cite{Boyd_etal1994:Book}. The obtained LMI is shown to be feasible if the system is observable. The feasibility is independent of selection of a homogeneity degree.
  \item The proposed LMI-based design covers also the case of a discontinuous  homogeneous observer
 being a version of  high order sliding mode (HOSM) algorithm \cite{Levant2003:IJC}, \cite{Jbara_etal2021:RNC}. To the best of authors knowledge, an LMI-based tuning of a HOSM observer has never been proposed before.
 \item  Finally, to  explain the basic idea of analysis of  the filtering homogeneous observer, we present a very simple LMI-based scheme of the so-called prescribed-time observer  design for linear MIMO system. This result is generalizes algorithms known in the prescribed-time control/observation theory \cite{Song_etal2017:Aut}, \cite{HollowayKrstic2019:TAC}, \cite{EfimovPolyakov2021:Book}, \cite{Zhou_etal2022:TAC}.\vspace{-1mm}
 \end{itemize}
The paper is organized as follows. First, we present a problem statement and provide some preliminary remarks about homogeneous systems. Next, the main result is presented and supported by numerical simulations.
At the end, some conclusions are given. All proofs are provided in the appendix.

\textit{Notation}:
$\R$ is the field of reals;  $\zero$ is the zero element of a vector space (e.g., $\zero \in \R^n$ means that $\zero$ is the zero vector); $\|\cdot\|$ is a norm in $\R^n$ (to be specified later);
a matrix norm for $A\in \R^{n\times n}$ is defined as $\|A\|=\sup_{x\neq \zero} \frac{\|Ax\|}{\|x\|}$;  $\lambda_{\min}(P)$ denote a minimal eigenvalue of a symmetric matrix $P=P^{\top}\in \R^{n\times n}$; $P\succ 0$ means that the symmetric matrix $P$ is positive definite; $C^1(\Omega_1,\Omega_2)$ denotes the set of continuously differentiable functions  $\Omega_1\subset \R^n\mapsto \Omega_2\subset \R^m$; $L^{\infty}(\R,\R^k)$ is the Lebesgue space of measurable uniformly essentially bounded functions $\R\mapsto \R^k$ with the norm defined by the essential supremum; $\|q\|_{L^{\infty}_{(t_0,t)}}=\mathrm{ess}\sup_{\tau\in (t_0,t)}\|q(\tau)\|$ for $q\in  L^{\infty}(\R,\R^k)$; we write  $\stackrel{a.e.}{=}$ (resp, $\stackrel{a.e.}{\leq}$) if the identity (resp., inequality) holds almost everywhere.

\section{Problem Statement}

Let us consider the quasi-linear  control system\vspace{-2mm}
\begin{equation}\label{eq:Ax+Bu,y=Cx}
	\left\{
	\begin{array}{l}
		\dot x=Ax + Bu +Eq(t,x,u),\\
		y=Cx,
	\end{array}
	\right. \quad t>0
\end{equation}
where $x(t)\in \R^n$ is the state variable, $y(t)\in \R^k$ is the measured output, $u : \R\mapsto \R^m$ is a known  (e.g., control) input, $A\in \R^{n\times n}$ is the system matrix, $B\in\R^{n\times m}$ is the matrix of input gains and the matrix $C\in\R^{k \times n}$ models the measurements of the state variables, $E\in \R^{n\times r}$ is  a known matrix  and $q:\R^{1+n+m}\mapsto \R^r$ is a possibly unknown bounded function:\vspace{-1mm}
\begin{equation}\label{eq:est_q}
|q(t,x,u)|\leq\bar q, \quad \forall t\geq 0, \quad \forall x\in \R^n, \quad \forall u\in \R^m.\vspace{-1mm}
\end{equation}
Our goal is to design a $\tilde \dn$-homogeneous observer  for the system \eqref{eq:Ax+Bu,y=Cx} in the sense of the following definition.
\begin{definition}
	A system \vspace{-2mm}
	\begin{equation}
		\dot x=f(t,x), \quad  t>0,\quad   f\in C(\R^n,\R^n)\vspace{-2mm}
	\end{equation}
	with the output measurements\vspace{-2mm}
	\begin{equation}
		y=h(x), \quad h\in C(\R^n,\R^k)\vspace{-2mm}
	\end{equation}
is said to be $\tilde{\dn}$-homogeneously observable of degree $\nu\in \R$ if there exists a system (called observer) of the form \vspace{-2mm}
	\begin{equation}\label{eq:nonlin_homogeneous_observer_Rn}
		\left\{
		\begin{array}{l}
			\dot z_1\!=\!f_1(t,z_1,z_2,y), \quad z_1(t)\in \R^n, \\
			\dot z_2\!=\!f_2(t,z_1,z_2,y), \quad  z_{2}(t)\in \R^{p},
		\end{array} \right. \vspace{-2mm}
	\end{equation}
 with $f_1 :\R^{1+n+p+k}\!\mapsto\! \R^n$, $f_2 :\R^{1+n+p+k}\! \mapsto \!\R^p$
	such that \vspace{-1mm}
	\begin{itemize}
		\item the error variable $\varepsilon=\left(\begin{smallmatrix} z_1-x \\ z_2\end{smallmatrix}\right)$ satisfies  a differential inclusion (or equation)\vspace{-2mm}
		\begin{equation}\label{eq:nonlin_error_observer}
			\dot \varepsilon\in G(\varepsilon),\quad G: \R^{n+p} \mapsto 2^{\R^{n+p}}, \vspace{-2mm}
		\end{equation}
		being  $\tilde \dn$-homogeneous  of degree $\nu\in \R$ (see below);
		\item
		the error system \eqref{eq:nonlin_error_observer} is
		globally uniformly asymptotically stable. \vspace{-1mm}
	\end{itemize}
\end{definition}
The representation \eqref{eq:nonlin_homogeneous_observer_Rn} covers observers of various structures.
Indeed, the case, $f_1=f(t,z_1,y)$ corresponds to the conventional observer, which variable $z_1$ must converge to the state $x$ of the system, while dependence of the observer on the internal variable $z_2$ can be utilized for the design of the so-called extended observers \cite{Li_etal2012:TIE}, prescribed-time observers \cite{HollowayKrstic2019:TAC} or filtering observers \cite{Levant2003:IJC}.

We solve the considered problem in the following way. First, we design a homogeneous  observer for the linear plant $q= \zero$ and provide a characterization of admissible observers parameters in the form of LMI. Next, we discover a sufficient condition to the matrix $E$  allowing the observer (designed for the linear model) to be applicable in the nonlinear case $q\neq \zero$ provided that $\bar q$ is small enough. We complete our investigations with the robustness analysis (in the ISS sense \cite{Sontag1989:TAC}) of the error system with respect to measurement noises and other unmodeled  additive perturbations of the plant \eqref{eq:Ax+Bu,y=Cx}.

\section{Generalized homogeneity}
\vspace{-1mm}
\subsection{Linear Dilation}\vspace{-2mm}
Let us recall that \textit{a family of  operators} $\dn(s):\R^n\mapsto \R^n$ with $s\in \R$ is  a \textit{group} if
$\dn(0)x\!=\!x$, $\dn(s) \dn(t) x\!=\!\dn(s+t)x$, $\forall x\!\in\!\R^n, \forall s,t\!\in\!\R$.
A \textit{group} $\dn$ is
a) \textit{continuous} if the mapping $s\mapsto \dn(s)x$ is continuous,  $\forall x\!\in\! \R^n$;
b) \textit{linear} if $\dn(s)$ is a linear mapping (i.e., $\dn(s)\in \R^{n\times n}$), $\forall s\in \R$;
c)  a \textit{dilation} in $\R^n$ if $\liminf\limits_{s\to +\infty}\|\dn(s)x\|=+\infty$ and $\limsup\limits_{s\to -\infty}\|\dn(s)x\|=0$,  $\forall x\neq \zero$.
Any linear continuous group in $\R^n$ admits the representation  \cite{Pazy1983:Book}\vspace{-2mm}
\begin{equation}\label{eq:Gd}
	\dn(s)=e^{sG_{\dn}}=\sum_{j=1}^{\infty}\tfrac{s^jG_{\dn}^j}{j!}, \quad s\in \R,\vspace{-2mm}
\end{equation}
where $G_{\dn}\in \R^{n\times n}$ is a generator of $\dn$. A continuous linear group \eqref{eq:Gd} is a dilation in $\R^n$ if and only if $G_{\dn}$ is an anti-Hurwitz matrix \cite{Polyakov2020:Book}. In this paper we deal only with continuous linear dilations.
A \textit{dilation} $\dn$ in $\R^n$ is
i) \textit{monotone} if the function $s\mapsto \|\dn(s)x\|$ is strictly increasing,  $\forall x\neq \zero$;
ii) 	\textit{strictly monotone} if  $\exists \beta\!>\!0$ such that $\|\dn(s)x\|\!\leq\! e^{\beta s}\|x\|$, $\forall s\!\leq\! 0$, $\forall x\in \R^n$.\\
The following result is the straightforward consequence of the existence of the quadratic Lyapunov function  for asymptotically stable LTI systems.\vspace{-2mm}
\begin{corollary}\label{cor:monotonicity}
	A linear continuous dilation in $\R^n$ is strictly monotone with respect to the weighted Euclidean norm $\|x\|=\sqrt{x^{\top} Px}$ with $0\prec P\in \R^{n\times n}$ if and only if
$	PG_{\dn}+G_{\dn}^{\top}P\succ 0,  P\succ 0.$\vspace{-1mm}
\end{corollary}
The standard dilation corresponds to $G_{\dn}=I_n$. The generator of the weighted dilation is $G_{\dn}\!=\!\diag(r_1,...,r_n)$.
\subsection{Canonical homogeneous norm}
Any linear  continuous and monotone  dilation in a normed vector space introduces also an alternative norm topology defined by the so-called canonical homogeneous norm \cite{Polyakov2020:Book}.
\begin{definition}
	\label{def:hom_norm_Rn}
	Let a linear dilation $\dn$ in $\R^n$ be  continuous and monotone with respect to a norm $\|\cdot\|$.
	A function $\|\cdot\|_{\dn} : \R^n \mapsto [0,+\infty)$ defined as  follows: $\|\zero\|_{\dn}=0$ and \vspace{-2mm}
	\begin{equation}\label{eq:hom_norm_Rn}
		\|x\|_{\dn}\!=\!e^{s_x}, \;  \text{where} \; s_x\in \R: \|\dn(-s_x)x\|\!=\!1, \; x\!\neq\! \zero
		\vspace{-2mm}
	\end{equation}
	is said to be a canonical $\dn$-homogeneous norm \index{canonical homogeneous norm} in  $\R^n$.
\end{definition}	
Notice that, by construction, $\|x\|_{\dn}=1\; \Leftrightarrow\;  \|x\|=1$. Due to the monotonicity of the dilation, $\|x\|_{\dn}< 1 \; \Leftrightarrow\;  \|x\|<1$ and $\|x\|_{\dn}>1 \; \Leftrightarrow\;  \|x\|>1$.

\begin{lemma}\cite{Polyakov2020:Book}\label{lem:hom_norm}
	If  a linear continuous dilation $\dn$ in $\R^n$  is  monotone with respect to a norm $\|\cdot\|$
	then 1)   $\|\cdot\|_{\dn} : \R^n\mapsto \R_+$ is single-valued and continuous on $\R^n$;
		2)		there exist $\sigma_1,\sigma_2
		\in \mathcal{K}_{\infty}$
		such that
		\begin{equation}\label{eq:rel_norm_and_hom_norm_Rn}
			\sigma_1(\|x\|_{\dn})\leq \|x\|\leq \sigma_2(\|x\|_{\dn}), \quad \quad
			\forall x\in \R^n;
		\end{equation}
		3)   $\|\cdot\|$ is locally Lipschitz continuous on $\R^{n}\backslash\{\zero\}$ provided that  $\dn$ is strictly monotone; 4) $\|\cdot\|_{\dn}$ is continuously differentiable on $\R^n\backslash\{\zero\}$ provided that $\|\cdot\|$ is continuously differentiable on $\R^n\backslash\{\zero\}$ and $\dn$ is strictly monotone.
\end{lemma}
For the $\dn$-homogeneous norm $\|x\|_{\dn}$ induced by the weighted Euclidean norm $\|x\|=\sqrt{x^{\top}Px}$ we have \cite{Polyakov2020:Book}\vspace{-2mm}
\begin{equation}\label{eq:hom_norm_derivative_Rn}
	\tfrac{\partial \|x\|_{\dn}}{\partial x}=\|x\|_{\dn}\tfrac{x^{\top}\dn^{\top}(-\ln \|x\|_{\dn})P\dn(-\ln \|x\|_{\dn})}{x^{\top}\dn^{\top}(-\ln \|x\|_{\dn})PG_{\dn}\dn(-\ln \|x\|_{\dn})x},\vspace{-2mm}
\end{equation}
\begin{equation}\label{eq:dilation_rates_in_Rn}
	\sigma_1(\rho)=\left\{
	\begin{smallmatrix}
		\rho^{\alpha} & \text{ if }  & \rho\leq 1,\\
		\rho^{\beta} & \text{ if} & \rho >1,
	\end{smallmatrix}
	\right. \quad \quad \sigma_2(\rho)=\left\{
	\begin{smallmatrix}
		\rho^{\beta} & \text{ if }  & \rho\leq 1,\\
		\rho^{\alpha} & \text{ if} & \rho >1,
	\end{smallmatrix}
	\right.
\end{equation}
where 	$\alpha=\frac{1}{2}\lambda_{\max}\left(P^{\frac{1}{2}} G_{\dn}P^{-\frac{1}{2}}+P^{-\frac{1}{2}}G^{\top}_{\dn} P^{\frac{1}{2}}\right)\geq$
$\beta=\frac{1}{2}\lambda_{\min}\left(P^{\frac{1}{2}} G_{\dn}P^{-\frac{1}{2}}+P^{-\frac{1}{2}}G^{\top}_{\dn} P^{\frac{1}{2}}\right)>0.$

\subsubsection{Homogeneous Functions and Vector fields}
Below we study systems that are symmetric with respect to a linear dilation $\dn$. The  dilation symmetry introduced by the following definition is known as a generalized  homogeneity \cite{Zubov1958:IVM}, \cite{Kawski1991:ACDS}, \cite{Rosier1992:SCL}, \cite{BhatBernstein2005:MCSS}, \cite{Polyakov2020:Book}.
\begin{definition}\cite{Kawski1991:ACDS}\label{def:hom_vf}
	A vector field $f:\R^n \mapsto \R^n$ is  $\dn$-homogeneous of
	degree $\mu\in \R$  if \vspace{-2mm}
	\begin{equation}\label{eq:homogeneous_operator_Rn}
		f(\dn(s)x)=e^{\mu s}\dn(s)f(x), \quad  \quad  \forall s\in\R, \quad \forall x\in \R^n.\vspace{-2mm}
	\end{equation}
\end{definition}
The  homogeneity of a mapping is inherited by other mathematical objects induced by this mapping.
In particular, solutions of $\dn$-homogeneous system\footnote{A system is homogeneous  if its is governed by a $\dn$-homogeneous vector field}
\begin{equation}\label{eq:nonlin_Rn}
	\dot x =f(x), \quad t\in \R, \quad x(t)\in \R^n,
\end{equation}
are symmetric with respect to the dilation $\dn$ in the following sense \cite{Zubov1958:IVM}, \cite{Kawski1991:ACDS}, \cite{BhatBernstein2005:MCSS}
\begin{equation}\label{eq:hom_solutions_Rn}
	x(t,\dn(s)x_0)=\dn(s)x(e^{\mu s}t,x_0),
\end{equation}
where $x(\cdot,z)$ denotes a solution of  \eqref{eq:nonlin_Rn}  with $x(0)=z\in \R^n$ and
$\mu\in \R$ is the homogeneity degree of $f$. The mentioned symmetry
of solutions implies many useful properties of homogeneous system such as equivalence of local and global results,  completeness for non-negative homogeneity degree, etc.
The system \eqref{eq:nonlin_Rn} always has strong solutions (at least in the sense of Filippov \cite{Filippov1988:Book}) if the  vector field $f$ is  locally bounded and measurable.
For the homogeneous system,  a uniform convergence of solutions to $\zero$  is equivalent to stability \cite{BhatBernstein2005:MCSS}.
\begin{theorem}\label{thm:hom_conv2stab}
	The $\dn$-homogeneous system \eqref{eq:nonlin_Rn} of degree $\mu<0$ is globally uniformly finite-time stable \cite{EfimovPolyakov2021:Book} if and only if
	any trajectory of the system reaches the origin in a finite time being locally bounded function of the initial state $x_0\in \R^n$.\vspace{-2mm}
\end{theorem}
We use the above theorem for analysis of the homogeneous observer. Its proof is given in Appendix.

\section{Filtering homogeneous observer}
\vspace{-1mm}
\subsection{Basic ideas of the observer design and analysis}
\vspace{-2mm}
Let us consider the following system\vspace{-2mm}
\begin{equation}\label{eq:ch14_dyn_hom_observer}
	\left\{\begin{array}{l}\dot z\!=\!Az\!+\!Bu\!+\!(L_0\!+\!|\xi|^{\nu-1}\dn(\ln |\xi|)L)(Cz\!-\!y), \\
		\dot \xi\!=\!-\rho|\xi|^{1+\nu}\sign(\xi),  \quad t\!>\!0,\quad \xi(0)\!=\!\xi_0\neq 0,
	\end{array}
	\right.\vspace{-2mm}
\end{equation}
where $z(t)\in \R^n$ is the observer state, $\xi(t)\in \R$ is an auxiliary variable, $\dn$ is a linear dilation in $\R^n$, $\nu\neq 0$,  $\gamma>0$ and $L_0, L\in \R^{n\times k}$ are parameters of the observer.
On the one hand, for $\xi=1$ then the first equation  in \eqref{eq:ch14_dyn_hom_observer} becomes the conventional
linear observer with the static gain $L_0+L$.
On the other hand, the differential equation for the variable $\xi$ has the explicit solution $\xi(t)=(|\xi_0|^{-\nu} +\nu \rho t)^{\frac{1}{-\nu}} \sign(\xi_0)$.  So the system \eqref{eq:ch14_dyn_hom_observer} can be interpreted as a linear observer  with a time-varying gain (dependent of $\xi(t)$).
However, considering the extended error equation \vspace{-2mm}
\begin{equation}\label{eq:ch14_dyn_hom_observer_error}
\!\;	\dot \varepsilon_{\xi}\!=\!g_{\xi}(\varepsilon_{\xi})\!:=\!\left(\!
	\begin{smallmatrix}
		A_0\varepsilon+|\xi|^{\nu-1}\dn(\ln |\xi|)LC\varepsilon\\
		-\rho |\xi|^{\nu+1}\sign(\xi)
	\end{smallmatrix}
	\!\right)\!,  A_0\!=\!A\!+\!L_0C\!\vspace{-2mm}
\end{equation}
for the \textit{augmented} error variable
$
\varepsilon_{\xi}=\left(\begin{smallmatrix}
	\varepsilon\\
	\xi
\end{smallmatrix}
\right),  \varepsilon=z-x,$ a homogeneity (dilation symmetry) can be established and utilized for the analysis of the error system. Indeed,  if the pair $(A,C)$ is observable, then  the linear gain $L_0$ and a linear dilation $\dn$ can be selected (see below) such that the matrix $A_0$ is $\dn$-homogeneous of degree $\nu$,\vspace{-2mm}
\begin{equation}\label{eq:hom_con_A_C}
	A_0\dn(s)\!=\!e^{\nu s}\dn(s)A_0 \; \text{ and } \; C\dn(s)\!=\!e^sC, \quad  \forall s\!\in\! \R. \vspace{-2mm}
\end{equation}
This implies  that the extended error equation is $\tilde \dn$-homogeneous of degree $\nu$ with respect to the dilation $\tilde \dn(s)=\left(\begin{smallmatrix}  \dn(s) & 0\\ 0 & e^s\end{smallmatrix}\right)$ in $\R^{n+1}$ as follows\vspace{-2mm}
\[
g_{\xi}(\tilde \dn(s)\varepsilon_{\xi})\!=\!e^{\nu s} \tilde \dn(s)g_{\xi}(\varepsilon_{\xi}), \quad \forall s\!\in\! \R, \forall  \varepsilon_{\xi}\!\in\! \R^{n+1}, \xi\!\neq\! 0.\vspace{-2mm}
\]
 In this case, for the variable
 $
 \epsilon=\dn(-\ln |\xi|)\varepsilon
 $
 we have
 $
 \dot \epsilon=-\frac{\dot \xi \sign(\xi)}{|\xi|} G_{\dn}\dn(-\ln |\xi|)\varepsilon+\dn(-\ln |\xi|)\dot \varepsilon
 $ and \vspace{-2mm}
 \begin{equation}\label{eq:epsilon_PT}
 \dot\epsilon=|\xi|^{\nu}(A_0+LC+\rho G_{\dn})\epsilon\vspace{-2mm}
 \end{equation}
 as long as $|\xi|\neq 0$.
 If the matrix $A_0+LC+\rho G_{\dn}$ is stable then $\epsilon$ is at least uniformly bounded for $t\geq 0$. Taking into account, the limit property of the dilation we derive that $\varepsilon=\dn(\ln |\xi|)\epsilon\to \zero$  as $\xi\to 0$. The only issue of the presented analysis is a proper selection
 of observer parameters $L_0,L$ and the generator of the dilation $G_{\dn}$.
\begin{theorem}\label{thm:ch14_dyn_hom_obs}
	Let $q=\zero$. For any observable  pair $\{A,C\}$ \vspace{-2mm}
	\begin{itemize}
		\item[\textbf{1)}]  the linear algebraic equation\vspace{-2mm}
		\begin{equation}\label{eq:ch14_Y_0G_0}
			G_0A-AG_0+Y_0C=A, \quad CG_0=\zero\vspace{-2mm}
		\end{equation}
		has a solution  $Y_0\!\in\! \R^{n\times m}$, $G_0\!\in\! \mathbb{R}^{n\times n}$  such that \vspace{1mm}
		\begin{itemize}
			\item[a)] the matrix \vspace{-1mm}
			\begin{equation}\label{eq:ch14_gen_Gd}
				G_{\mathbf{d}}\!=\!I_n\!+\!\nu G_0\vspace{0mm}
			\end{equation} is anti-Hurwitz for $\nu\!\geq -1/\tilde n$, where $\tilde n\!\in\! \mathbb{N}$ is a minimal number such that
			$\rank	\left(\begin{smallmatrix}
					C\\
					CA\\
					...\\
					CA^{\tilde n-1}
				\end{smallmatrix}
				\right)=n;$

			\item[b)] the matrix $I_n+G_0$ is invertible and  the matrix \vspace{-2mm}
			\begin{equation}
				A_0=A-(I_n+G_0)^{-1}Y_0C\vspace{-2mm}
			\end{equation}
			satisfies	the identity \vspace{-2mm}
			\begin{equation}\label{eq:ch14_hom_A0}\vspace{-2mm}
				A_0G_{\dn}=(G_{\dn}+\nu I_n)A_0 \quad \text{ and } \quad CG_{\dn}=C;\vspace{0mm}
			\end{equation} 					
		\end{itemize}
		
		\item[\textbf{2)}]  the following LMI\vspace{-2mm}
		\begin{equation}\label{eq:ch14_LMI_obs_Rn}
		\!\!\!\!\!\!\!	\begin{array}{c}			PA_0\!+\!A_0^{\top}P+\!YC\!+\!C^{\top}Y^{\top}\!\!+\!\rho(PG_{\dn}+\!G_{\dn}^{\top}P)\!\preceq\! 0, \\ PG_{\dn}+G_{\dn}^{\top}P\succ 0, \quad P=P^{\top}\succ 0
			\end{array}			\vspace{-2mm}
		\end{equation}
		has a solution $P\in \mathbb{R}^{n\times n}$, $Y\in \mathbb{R}^{n\times k}$ for any $\rho>0$;
		
		\item[\textbf{3)}] the error equation \eqref{eq:ch14_dyn_hom_observer_error}  is
		$\tilde \dn$-homogeneous of degree $\nu$ with respect to the dilation
	$
		\tilde \dn(s)=\left(
		\begin{smallmatrix}
			\dn(s) & \zero\\
			\zero & e^s
		\end{smallmatrix}
		\right)
		$
		in $\R^{n+1}$, 		where $\dn(s)=e^{sG_{\dn}}$ is a dilation in $\R^n$ and  $s\in \R$;

		\item[\textbf{4)}] the  system \eqref{eq:ch14_dyn_hom_observer} with $L_0=-(I_n+G_{0})^{-1}Y_0$ and $L=P^{-1}Y$ has a solution defined on $[0, T_0)$, where
		 $T_0=+\infty$ for $\nu\!\geq\! 0$ and  $T_0\!=\!\frac{\xi^{-\nu}_0}{-\nu\alpha}\!<\!+\infty$ for $\nu\!<\!0$; moreover,  for any  $\xi_0\neq 0$ the  error variable $\varepsilon_{\xi}$ converges to $\zero$\vspace{-2mm}
		$$
		\varepsilon_\xi(t) \to \zero \quad \text{ as } \quad t\to T_0\vspace{-2mm}		$$
		in the fixed time $T_0<+\infty$ if $\nu<0$ or asymptotically $T_0=+\infty$ if $\nu\geq 0$.
	\end{itemize}
\end{theorem}
This theorem is given without proof since the first three conclusions of the above theorem follows from the Kalman duality: \textit{if the pair $\{A,C\}$ is observable  then the pair $\{A^{\top},C^{\top}\}$ is controllable}. In this case, denoting $\tilde A=A^{\top}$, $\tilde B=C^{\top}$, $\tilde G_0=-G_0^{\top}$,  $\tilde Y_0= Y_0^{\top}$ and $\tilde Y=Y^{\top}$, we reduce the feasibility analysis of the algebraic systems \eqref{eq:ch14_Y_0G_0} and \eqref{eq:ch14_LMI_obs_Rn} to the same analysis already done for the homogeneous controller design (see, \cite{Polyakov_etal2016:RNC}, \cite{Zimenko_etal2020:TAC}, \cite{Polyakov2020:Book}). The final conclusion is the straightforward consequence of \eqref{eq:epsilon_PT} and \eqref{eq:ch14_LMI_obs_Rn}.

For $\nu<0$ we derive $\xi(T_0)=0$ and $\epsilon(T_0)=\zero$, i.e., in the case of the negative homogeneity degree, the observer \eqref{eq:ch14_dyn_hom_observer_error} is the so-called prescribed-time observer \cite{HollowayKrstic2019:TAC}.  The critical disadvantage  of
the prescribed-time observer with the time-varying gain is its high sensitivity with respect to the measurement noise. Indeed, if the output measurements are noised $y=Cx+\eta$, then, independently of the estimation error, the noise $\eta$ will be amplified with the gain $\xi^{\nu-1}\dn(-\ln |\xi|)L$, which tends to infinity as $t\to T_0$. This drastically degrades the precision of prescribed-time  systems \cite{Aldana-Lopeza_etal2023:Aut}. A possible way to eliminate this drawback is to make the variable $\xi$ to be dependent on the available component of the estimation error  $y-Cz=C\varepsilon$. The above analysis is  based on the change of the variable $\epsilon=\dn(-\ln |\xi|)\varepsilon$, which is well-defined only for $|\xi|\neq 0$. Below we show that the similar analysis remains possible for the filtering homogeneous observer.

\vspace{-1mm}
\subsection{Main results}
\vspace{-2mm}
Let us consider the system \eqref{eq:Ax+Bu,y=Cx} with $q\equiv\zero$.
Inspired by \cite{Nekhoroshikh_etal2022:CDC}, let us define the observer as follows\vspace{-2mm}
\begin{equation}\label{eq:ch14_ext_hom_observer}
	\left\{\begin{array}{l}\dot z=Az+Bu+L_0 (Cz-y)+\sigma^{\nu-1}\dn(\ln \sigma)L\psi\\
		\dot \psi= Cz-y+\sigma^{\nu}\tilde L \psi,\\
		\dot \xi=	\left|		\psi^{\top}\!\left(\!Cz\!-\!y\!+\!\sigma^{\nu}\tilde L\psi\right)\right|\!-\!\gamma \! \!\left\lfloor\xi\!-\!\frac{|\psi|^2}{2}\right\rceil^{\frac{\nu}{2}+1}\!,  \\
		\sigma=\sqrt{\max\left\{\xi, \frac{|\psi|^2}{2}\right\}},
	\end{array}
	\right.\vspace{-2mm}
\end{equation}
with $z(0)=z_0\in\R^n, \psi(0)\!=\!\psi_0\in\R^k,  \xi(0)\!=\!\xi_0\in\R$ and $t\geq 0$,
where $A,B,C,L_0,L,\xi$ are as before, $\psi(t)\in \R^k$ is the state of the output filter, $|\psi|^2=\psi^{\top}\psi$, $\tilde L\in \R^{k\times k}$, $\rho>0$ and  the linear dilation $\dn$ is defined below.
As before, the variable $\xi$ specifies a homogeneous scaling of the observer gains, but this variable also depends on the state of the output filter with the state $\psi$.  The gain tends to infinity only if the error tends to zero. This dependence prevents the infinite grow of the gain  $\sigma^{\nu-1}\dn(\ln \sigma)L$ in the case of noised measurement, while the filter itself improves the observation quality \cite{Jbara_etal2021:RNC}, \cite{Nekhoroshikh_etal2022:CDC}.

For $q=\zero$ the extended error equation has the form \vspace{-2mm}
\begin{equation}\label{eq:ch14_ext_hom_observer_error}
	\dot \varepsilon_{\sigma}\!=\!g(\varepsilon_{\sigma})\!:=\!
	\left(
	\begin{smallmatrix}
		C\varepsilon+|\sigma|^{\nu}\tilde L \psi\\
		A_0\varepsilon+|\sigma|^{\nu-1}\dn(\ln |\sigma|)L\psi\\	
		\left|
		\psi^{\top}\!\left(C\varepsilon+\sigma^{\nu}\tilde L\psi\right)\right|-\gamma \left\lfloor\xi\!-\!\frac{|\psi|^2}{2}\right\rceil^{\frac{\nu}{2}+1}\!
	\end{smallmatrix}
	\right)\!, \vspace{-2mm}
\end{equation}
where $A_0=A+L_0C$ and
$\varepsilon_{\sigma}=\left(
\begin{smallmatrix}
	\psi\\
	\varepsilon\\
	\xi
\end{smallmatrix}
\right),  \varepsilon=z-x.$
Below we show that, the variable $\sigma$ remains positive as long as  the norm of the  error is nonzero. This substantially simplifies the analysis of the error equation allowing a non-conservative algebraic conditions for tuning of the observer parameters to be obtained.
\begin{theorem}\label{thm:ch14_ext_hom_observer}
Let $q=\zero$, the pair $\{A,C\}$ be observable and the parameters $G_{0}\in \R^{n\times n}, \tilde n\in \N$, $L_0\in \R^{n\times k}$  and $A_{0}\in \R^{n\times n}$ be defined as in Theorem \ref{thm:ch14_dyn_hom_obs}. If
\vspace{-2mm}
	\begin{equation}
		G_{\dn}=I_n+\nu(G_{0}+I_n), \quad \dn(s)=e^{s G_{\dn}}, s\in \R, \vspace{-2mm}
	\end{equation}
	\begin{equation}
		\bar G=
		\left(
		\begin{smallmatrix}
			I_k & \zero\\
			\zero &G_{\dn}
		\end{smallmatrix}
		\right),  \quad
				\bar A_0=      \left(
		\begin{smallmatrix}
			\zero & C\\
			\zero&  A_0
		\end{smallmatrix}
		\right),  \quad \bar C=(I_k \;\;\zero )
	\end{equation}
 then the LMI \vspace{-2mm}
 		\begin{equation}\label{eq:ch14_LMI_ext_obs_Rn}
 	\begin{array}{c}
 		\bar P\bar A_0\!+\!\bar A_0^{\top}\bar P+\!\bar Y\bar C\!+\!\bar C^{\top}\bar Y^{\top}\!\!+\!\rho(\bar P\bar G+\!\bar  G^{\top}\bar P)\!\preceq\!0, \\ \bar P\bar G+\bar G^{\top}\bar P\succ 0, \quad \bar P=\bar P^{\top}\succ 0
 	\end{array}			
 \end{equation}
is feasible with respect to $\bar P\in \R^{(k+n)\times (k+n)}$ and $\bar Y\in \R^{(k+n)\times k}$ for any $\rho>0$. The error  equation \eqref{eq:ch14_ext_hom_observer_error} is \vspace{-2mm}
	\begin{itemize}
		\item[1)] $\dn_{\sigma}$-homogeneous of degree $\nu$ with respect to the dilation \vspace{-2mm}
		\begin{equation}
			\dn_{\sigma}(s)=\left(
			\begin{smallmatrix}
				e^{s \bar G} &
				\zero\\
				\zero & e^{2s}
			\end{smallmatrix}
			\right), \quad s\in \R;\vspace{0mm}
		\end{equation}
		\item[2)] continuous (i.e., $g_{\sigma}\in C(\R^{k+n+1},\R^{k+n+1})$) for $\nu>-\frac{1}{\tilde n+1}$ and piece-wise continuous if $\nu=-\frac{1}{\tilde n+1}$;
		\item[3)] globally finite-time stable for $\nu\in\left[-\frac{1}{\tilde n+1 },0\right)$ provided that
		the observer gains are defined as follows\vspace{-2mm}
		\begin{equation}
			\left(
			\begin{smallmatrix}
				\tilde L\\
				L	
			\end{smallmatrix}
			\right)=\bar L:=\bar P^{-1}\bar Y,\vspace{-2mm}
		\end{equation}
		where the pair $(\bar P,\bar Y)$ is a solution of \eqref{eq:ch14_LMI_ext_obs_Rn}	with  $\rho>\frac{\gamma}{2}$.
	\end{itemize}
\end{theorem}
The observability of the pair $\{A,C\}$ implies the observability of the pair $\{\bar A_0,\bar C\}$ and the feasibility
of the LMI \eqref{eq:ch14_LMI_ext_obs_Rn}. The above LMI is non-conservative in the sense that the observability of the pair $\{A,C\}$  is the necessary and sufficient condition for its feasibility.

 For the integrator chain
$A=\left(\begin{smallmatrix} \zero & I_{n-1}\\0 & \zero \end{smallmatrix}\right), C=(1\;\;0 \;\;...\;\;0)$, the observer \eqref{eq:ch14_ext_hom_observer} is similar to the observer given in \cite{Jbara_etal2021:RNC} with the filter of the order 1. The only difference is the use of the internal variable $\xi$, which controls the observer gain. Both observers simply coincide for $\xi\equiv \zero$.
Moreover, the case $\nu=-\frac{1}{\tilde n+1}$ corresponds to a discontinuous HOSM  observer.  The HOSM observers  are known to be applicable to the system \eqref{eq:Ax+Bu,y=Cx} with unknown bounded input/nonlinearity $q$ \cite{Levant2003:IJC}, \cite{Moreno2023:RNC}. However, this is possible only under special restrictions to the matrix $E$. The following corollary provides the corresponding condition for the  filtering homogeneous  observer.

\begin{corollary}\label{cor:HOSM}
	Let the filtering homogeneous observer \eqref{eq:ch14_ext_hom_observer} be designed as in
	Theorem \ref{thm:ch14_ext_hom_observer} for $\nu=-\frac{1}{\tilde n+1}$. If \vspace{-2mm}
	\begin{equation}\label{eq:rest_E}
	(\tilde n+1)G_{\dn}E=E\vspace{-2mm}
	\end{equation}
	and $q$ satisfies \eqref{eq:est_q}, then  the error equation
	$
		\dot \varepsilon_{\sigma}\!=\!g(\varepsilon_{\sigma})+\left(\begin{smallmatrix} \zero\\ E\\
		0\end{smallmatrix}\right)q
$
	is globally uniformly finite-time stable provided that
	$\bar q>0$  is small enough.
\end{corollary}
For the integrator chain discussed above, the dilation has the diagonal generator $G_{\dn}=\diag(1+\nu ,1+2\nu...,1+ \nu n )$ and the identity \eqref{eq:rest_E} holds for $E=(0,...,1)^{\top}\in \R^n$, since $\tilde n=n$. This perfectly fits the results of \cite{Levant2003:IJC}, \cite{Moreno2023:RNC}.

\subsection{Robustness analysis}
Let  us assume that the measurements are noised and the system is perturbed \vspace{-2mm}
\begin{equation}\label{eq:Ax+Bu,y=Cx_noised}
	\left\{
	\begin{array}{l}
		\dot x=Ax + Bu +q_x(t),\\
		y=Cx+q_y(t),
	\end{array}
	\right. \quad t>t_0\vspace{-2mm}
\end{equation}
where $x,A,B,C$ are as before and $q_x\in L^{\infty}(\R,\R^n)$ is additive perturbation and $q_y\in L^{\infty}(\R,\R^k)$ is measurement noise. In this case, the error equation has the form\vspace{-1mm}
\begin{equation}\label{eq:ch14_ext_hom_observer_error_error}
\!\,	\dot \varepsilon_{\sigma}\!=\!\bar g(\varepsilon_{\sigma}, \tilde q)\!:=\!\!
	\left(\!
	\begin{smallmatrix}
		C\varepsilon-q_y+|\sigma|^{\nu}\tilde L \psi\\
		A_0\varepsilon+|\sigma|^{\nu-1}\dn(\ln |\sigma|)L\psi-q_x\\	
		\left|
		\psi^{\!\top}\!\left(C\varepsilon-q_y+\sigma^{\nu}\tilde L\psi\right)\right|-\gamma
		 \left\lfloor\xi\!-\!\frac{|\psi|^2}{2}\right\rceil^{\frac{\nu}{2}+1}
	\end{smallmatrix}\!
	\right)\!, \!\!\!\vspace{0mm}
\end{equation}
where $\tilde q=(q_x^{\top},q_y^{\top})^{\top}$. Recall \cite{SontagWang1996:SCL} that the error system is Input-to-State Stable (ISS) with respect to $\tilde q$ if there exist $\beta\in \mathcal{KL}$, $\gamma\in \mathcal{K}$ such that \vspace{-2mm}
\[
|\varepsilon_{\sigma}(t)|\leq \beta(|\varepsilon_{\sigma}(t_0)|,t-t_0)+\gamma (\|\tilde q\|_{L^{\infty}_{(t_0,t)}}), \quad \forall t\geq t_0,\vspace{-2mm}
\]
for any $\varepsilon_{\sigma}(t_0)\in \R^{n+k+1}$ and any $\tilde q\in L^{\infty}(\R,\R^{n+k})$.
\begin{corollary}
	The error equation \eqref{eq:ch14_ext_hom_observer_error_error} is ISS with respect to $\tilde q$ if $\nu>-\frac{1}{\tilde n+1}$ and ISS with respect to $q_y$ if $\nu=-\frac{1}{\tilde n+1}$.\vspace{-2mm}
\end{corollary}
This corollary is given without proof since the result trivially follows from \cite{Ryan1995:SCL}, \cite{Andrieu_etal2008:SIAM_JCO}, \cite{Polyakov2020:Book}, where  it is proven that the dilation symmetry of the perturbed error system: $\bar g(\dn_{\sigma}(s)\varepsilon_{\sigma}, \dn_q(s)\tilde q)=e^{\nu s}\dn_{\sigma}(s) \bar g(\varepsilon_{\sigma},\tilde q),\forall s\!\in\! \R$ with $\dn_q(s)\!=\!\diag \{e^{\nu s}\dn(s),e^s I_k\}$
and an asymptotic stability of the unperturbed error \eqref{eq:ch14_ext_hom_observer_error} imply ISS provided that $\dn_{q}$ is a dilation in $\R^{n+k}$. For $\nu=-\frac{1}{\tilde n+1}$ the group $e^{\nu s}\dn(s)$ is not a dilation, so ISS can be guaranteed only with respect to $q_y$ and by Corollary \ref{cor:HOSM}, the additive perturbation
$q_x=Eq$ can be completely rejected in this case provided that $\bar q$ is small enough.

\section{Numerical example}
Let us consider a linearization (in the upper unstable position) of the rotary inverted pendulum  \cite{Lopez-Ramirez_etal2018:Aut}:
\[
A=\left(
\begin{smallmatrix}
      0     &    0 &   1     &   0\\
         0    &     0  &       0  &  1\\
         0 & 152.0057 & -12.2542 &  -0.5005\\
         0 & 264.3080  &-12.1117 &  -0.8702  \\
\end{smallmatrix}\right),\quad
B=\left(
\begin{smallmatrix}
         0\\
         0\\
   50.6372\\
   50.0484\\
\end{smallmatrix}\right)
\]
and $C=\left(\begin{smallmatrix} 1 & 0 & 0 & 0\\0 & 1 & 0 & 0\end{smallmatrix}\right)$. Taking $\nu=-1/3$ we design a HOSM-based filtering homogeneous observer.  The homogenization gain $L_0$ and the dilation matrix  we find by solving the linear algebraic equation \eqref{eq:ch14_Y_0G_0}:
\[
 L_0\!=\!\left(\!
 \begin{smallmatrix}
   12.2542 &   0.5005\\
   12.1117 &   0.8702\\
 -156.227 &-158.574\\
 -158.959 &-271.127
 \end{smallmatrix}\!
 \right)\!, G_{\dn}\!=\!\left(\!
 \begin{smallmatrix}
    2/3 &       0   &      0   &      0\\
         0  &  2/3  &       0   &      0\\
   -4.0847  & -0.1668 &  1/3   &      0\\
   -4.0372  & -0.2901 &        0  &  1/3
   \end{smallmatrix}\!
 \right)\!.
\]
The parameters $\tilde L=\left(\begin{smallmatrix}
      -13.9248  &  0.0000\\
         0  & -13.9248\\
\end{smallmatrix}\right)$ and $L=\left(\begin{smallmatrix}
    -90.01293   &  0\\
    0      &   -90.01293\\
          969.1348    &     45.0500\\
          1090.210    &     -55.5684
\end{smallmatrix}\right)$ are selected by solving the LMI \eqref{eq:ch14_LMI_ext_obs_Rn} for $\rho=3/2$, so the parameter $\gamma=2.75$ satisfies the inequality $\rho>\gamma/2$.

For comparison reasons we consider the linear (Luenberger) observer
\cite{Luenberger1964:TMI} $\dot z^{\rm lin}=Az^{\rm lin} +Bu+L^{\rm lin}(Cz^{\rm lin}-y)$ with the gain $L^{\rm lin}=\left(
\begin{smallmatrix}
  -10.9008 &   0.5005\\
   12.1117 & -22.0156\\
   34.0122 &-147.1205\\
  121.4870 &-343.9178\\
\end{smallmatrix}
\right)$.

\begin{figure}[h]
\centering
\includegraphics[width=.23\textwidth]{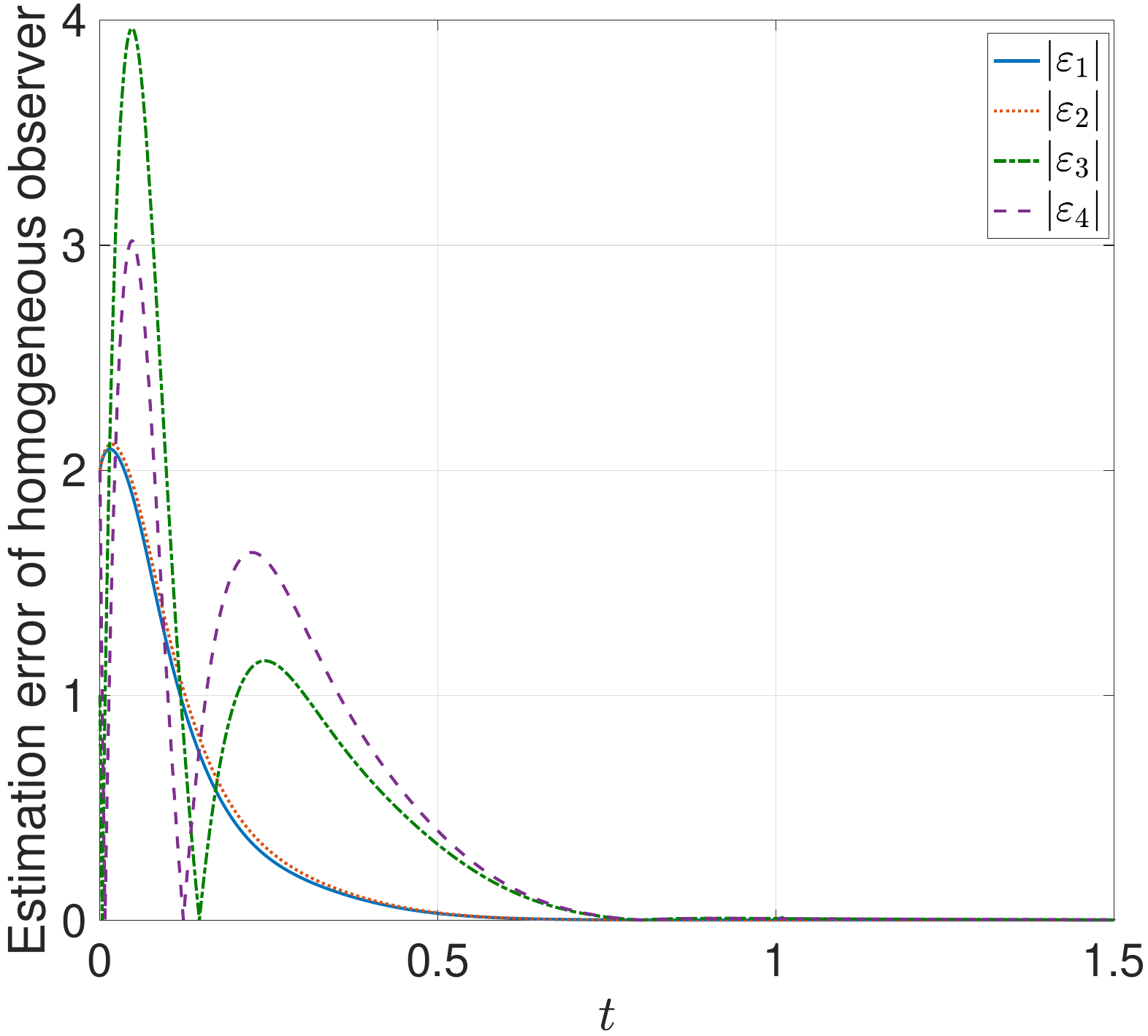}
\includegraphics[width=.23\textwidth]{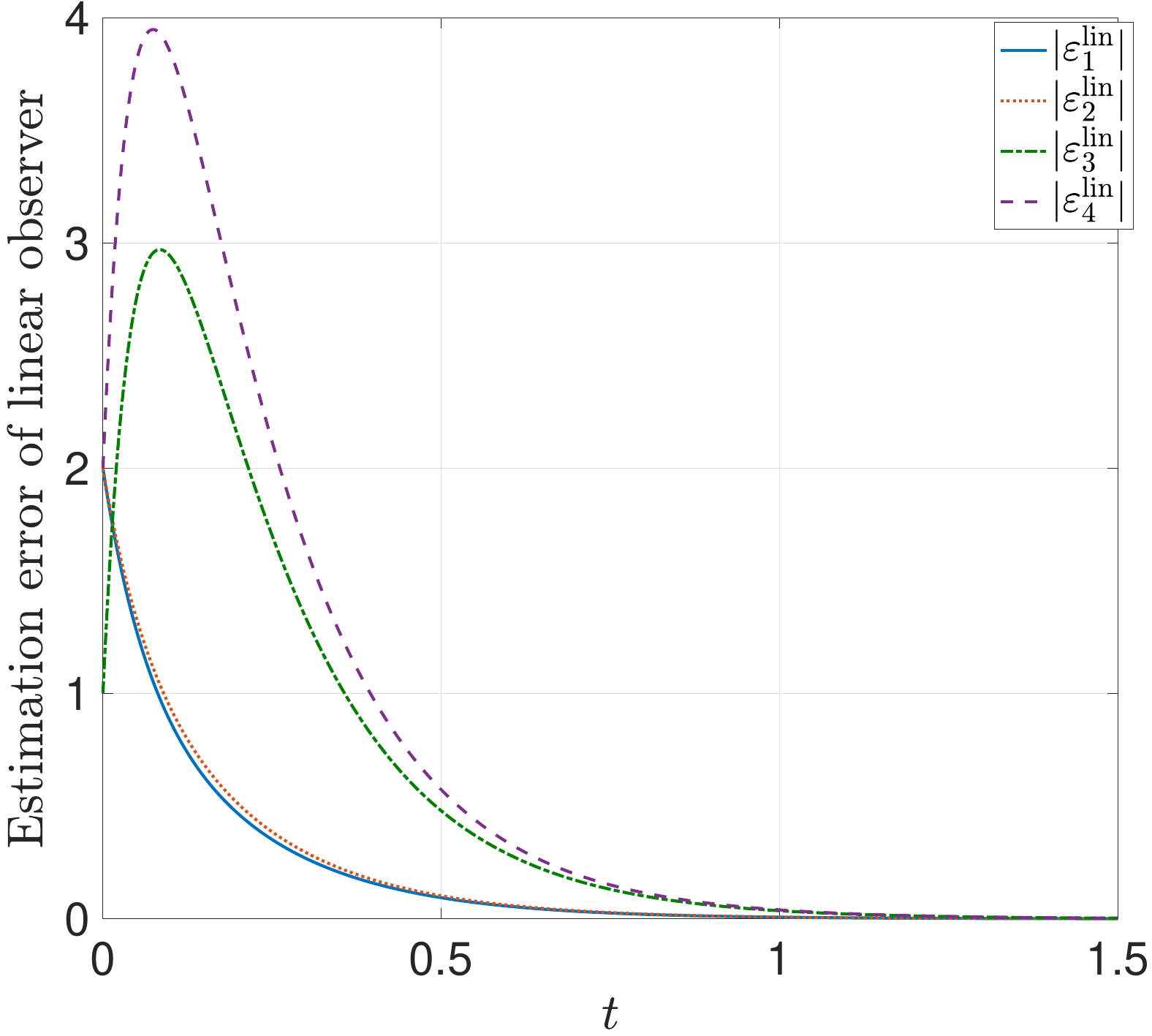}
\caption{Estimation errors for homogeneous (left) and linear (right) observers in the nominal case}\label{fig:nom}
\end{figure}

For simulation we select $x(0)=(2,2,1,2)^{\top}$, we initiate the  states of all observers at zero and we define the control input as $u=(2,-35,1.5,-3)z$, where $z$ is the component of the state of the filtering homogeneous observer. For simulations we use the explicit Euler method with the step size $h=0.5\cdot 10^{-4}$. The estimation errors $\varepsilon=z-x$ and $\varepsilon^{\rm lin}=z^{\rm lin}-x$ for the homogeneous and linear observers respectivly are presented on Fig. \ref{fig:nom}. Both observers show a convergence to the real state of the system. The estimation error at the time $t=1.5$ is about $\|x(1.5)-z(1.5)\|\approx \|z^{\rm lin}(1.5)-x(1.5)\|\approx 0.4 \cdot 10^{-4}$. Due to numerical error of the Euler method, the exact convergence to the system state cannot be obtained.

\begin{figure}[h]
\centering
\includegraphics[width=.23\textwidth]{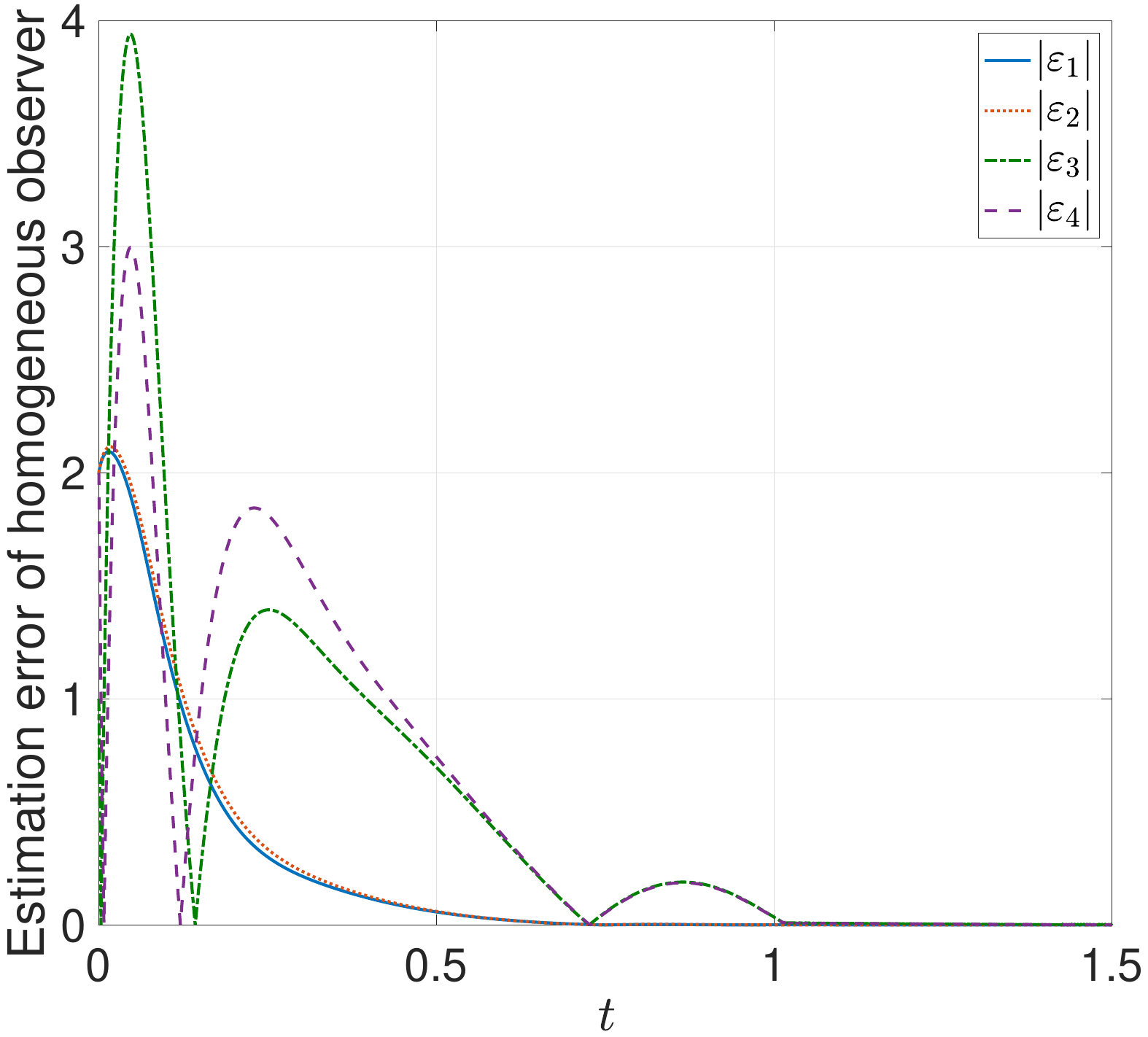}
\includegraphics[width=.23\textwidth]{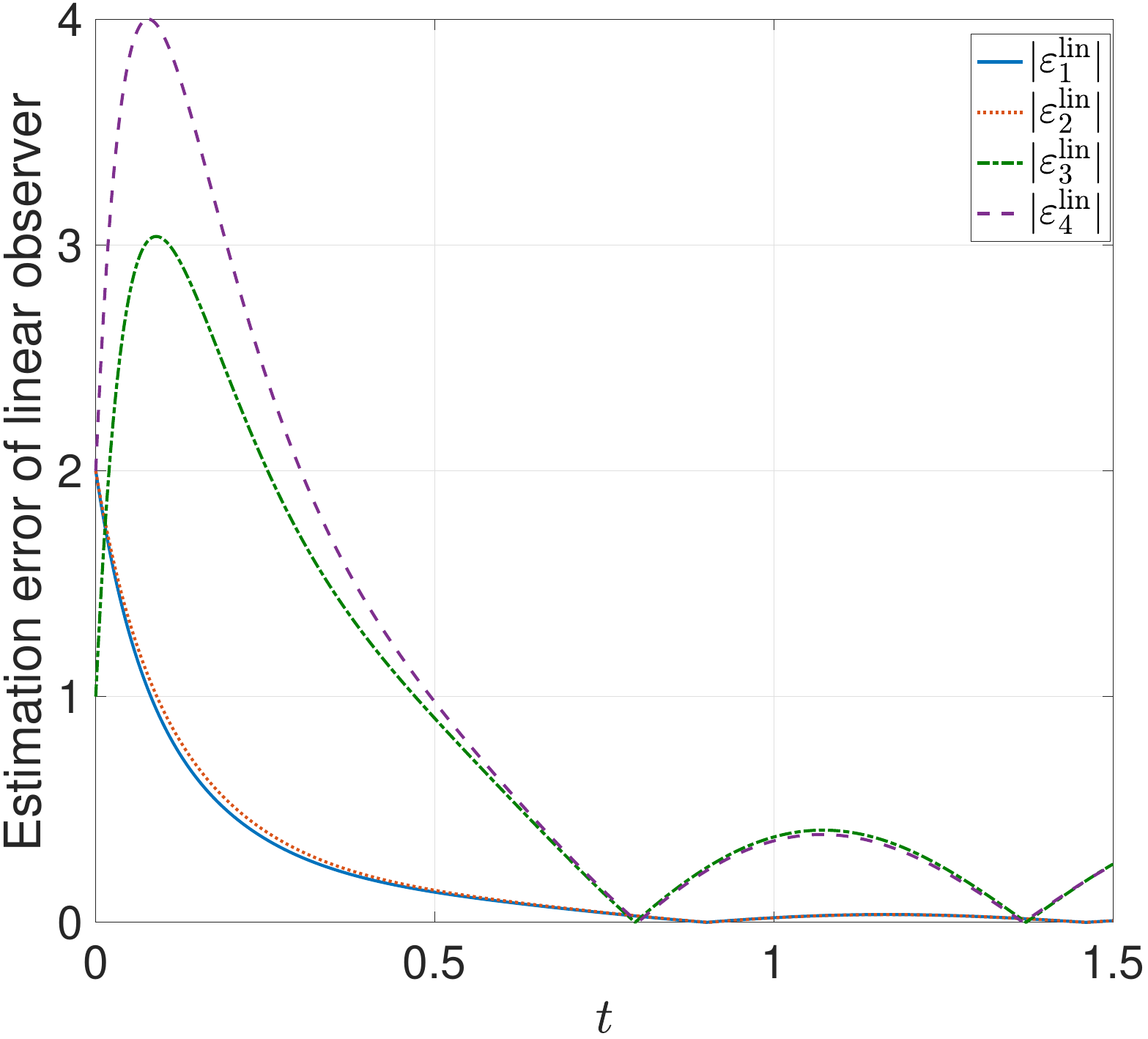}
\caption{Estimation errors for homogeneous (left) and linear (right) observers with the additive perturbation $0.1B\sin(5t)$}\label{fig:pert}
\end{figure}

\begin{figure}[h]
\centering
\includegraphics[width=.23\textwidth]{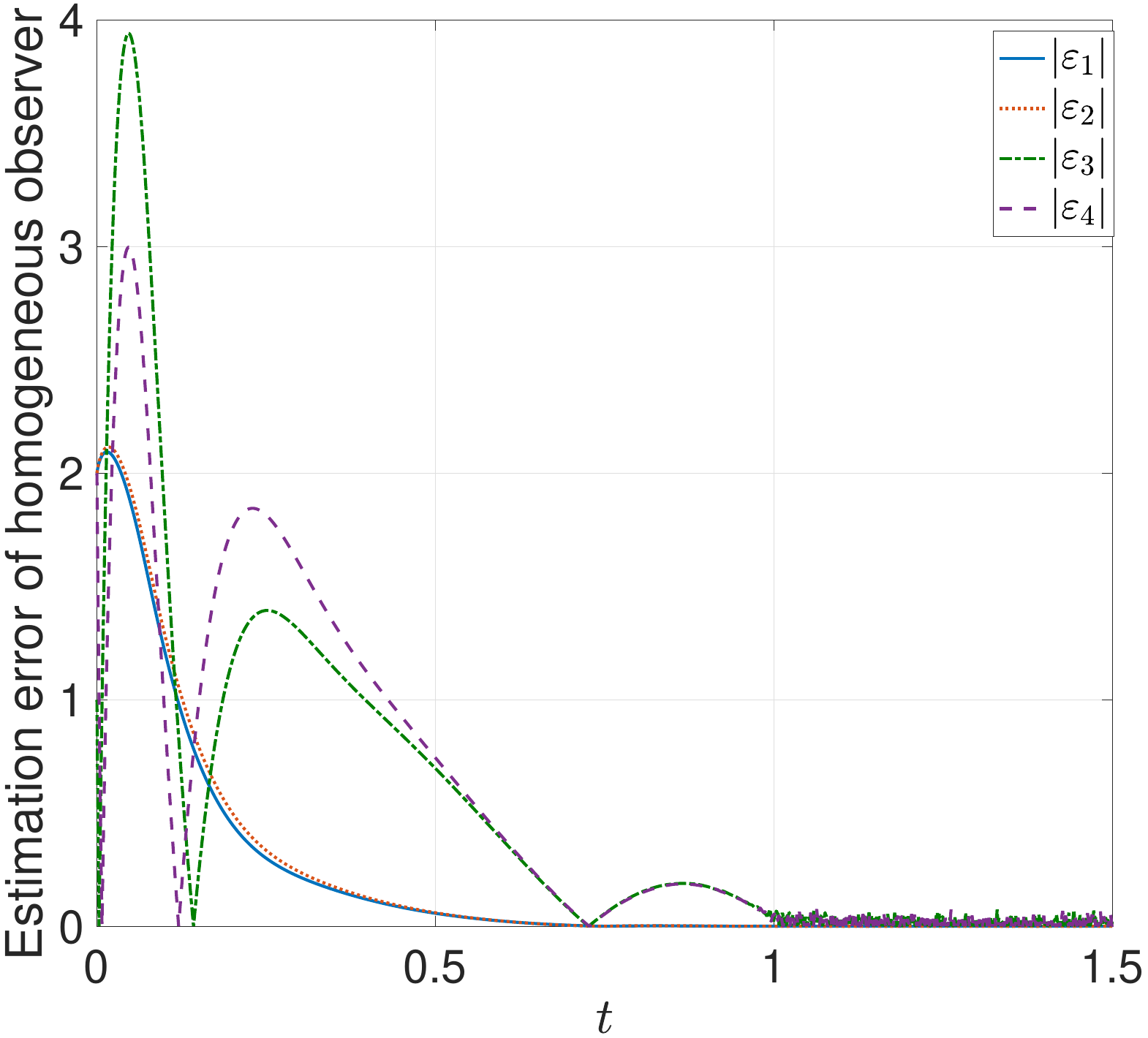}
\includegraphics[width=.23\textwidth]{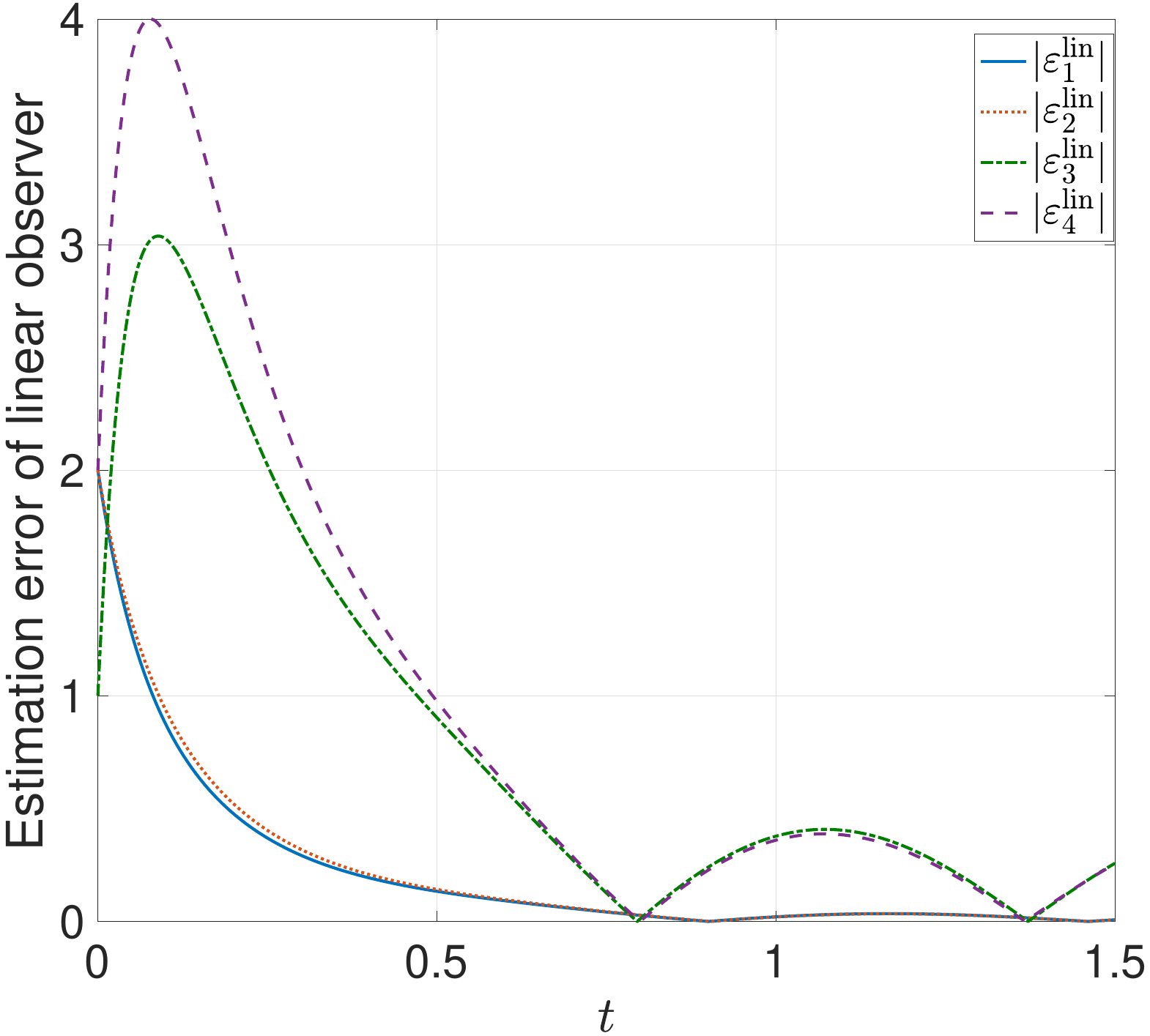}
\caption{Estimation errors for homogeneous (left) and linear (right) observers with the additive perturbation $0.1B\sin(5t)$ and with the noisy measurements of the magnitude $0.001$}\label{fig:noisy}
\end{figure}

By Corollary \ref{cor:HOSM} the HOSM-based homogeneous observer is able to reject unknown additive perturbation $Eg$ of some magnitude, provided that the matrix $E$ satisfies the identity \eqref{eq:rest_E}, which in particular holds for $E=B$. The simulation results for $g(t)=0.1\sin(5t)$ are depicted at Fig. \ref{fig:pert}.
To demonstrate the robustness (ISS of the homogeneous observer with respect to the measurement noises), the Fig. \ref{fig:noisy} presents the simulation results for the case of the matched additive perturbation considered above and a uniformly distributed measurement noise of the magnitude $0.001$. In this scenario the proposed homogeneous observer is clearly outperforming the linear counterpart.
\section{Conclusions}
The filtering homogeneous observer for MIMO system is developed. Its tuning is based on solving of an LMI, which is always feasible in the case of observable system. An analog of filtering HOSM observer is designed as well. It is shown that this observer is robust (non-sensitive) with respect to some ``matched'' unknown signals (or system nonlinearities) of some magnitude. This feature is well-known for HOSM differentiators \cite{Levant2003:IJC} and observers \cite{Moreno2023:RNC}.

\section{Appendix}
\subsection{The proof of Theorem \ref{thm:hom_conv2stab}}
	The \textit{Necessity} follows from the definition of the uniform finite-time stability. \textit{Sufficiency.} Let $T(r)>0$ be a supremum of reaching times over all initial conditions $x_0$ belonging to the homogeneous sphere
	of the radius $r>0$ (i.e.,  $\|x_0\|_{\dn}=r$).
	Since the reaching time is a locally bounded function of the initial state $x_0$ then $T(r)<+\infty$ for any finite $r>0$.
	Moreover,  the $\dn$-homogeneity of the solutions (see the formula \eqref{eq:hom_solutions_Rn}) implies that \vspace{-2mm}
	\[
	T(e^{s}r)=e^{-\mu s} T(r), \quad \forall s\in \R, \quad \forall r>0.\vspace{-2mm}
	\]
	The latter means that $T(r)\to 0$ as $r\to \zero$, i.e., $T$ is continuous at zero.
	
	Let us show that the system is globally Lyapunov stable.
	Suppose the contrary,  i.e., $\forall \delta\in (0,1),
	\exists x_0\in \R^n : \|x_0\|_{\dn}= \delta$ and $\exists t^*=t^*(\delta)>0 : \|x(t^*,x_0)\|_{\dn}=1$,
	and $\delta<\|x(t)\|_{\dn}<1$ for all $t\in (0, t^*)$,
	where $\|\cdot\|_{\dn}$ is a $\dn$-homogeneous norm induced by the weighted dilation $\|x\|=\sqrt{x^{\top}Px}$.	
	
	Using the formula \eqref{eq:hom_norm_derivative_Rn} we derive \vspace{-2mm}
	\[
	\frac{d}{d t} \|x\|_{\dn}\leq \|x\|_{\dn} \tfrac{x^{\top}\dn^{\top}(-\ln \|x\|_{\dn})P\dn(-\ln \|x\|_{\dn})f(x)}{x^{\top} \dn^{\top}(-\ln \|x\|_{\dn})PG_{\dn}\dn(-\ln \|x\|_{\dn})x}.\vspace{-2mm}
	\]
	Applying $\dn$-homogeneity of $f$ we obtain\vspace{-2mm}
	\[
		\frac{d}{d t} \|x\|_{\dn}=
 \|x\|_{\dn}^{1+\mu} \tfrac{x^{\top}\dn^{\top}(-\ln \|x\|_{\dn})Pf(\dn(-\ln \|x\|_{\dn})x)}{x^{\top} \dn^{\top}(-\ln \|x\|_{\dn})PG_{\dn}\dn(-\ln \|x\|_{\dn})x}.\vspace{-2mm}
	\]

	Hence, taking	$M=\sup_{\|y\|=1} \|f(y)\|<+\infty$ we derive
	$\|x_{x_0}(t)\|_{\dn}^{-\mu}\leq \|x_0\|_{\dn}^{-\mu}+\beta M t$ for all $t\geq 0$ and
	$
	\|x_{x_0}(t)\|_{\dn}<1 \text{ for all } t\in \left(0, \frac{1-\delta^{-\mu}}{M\beta}\right),
	$
where $\beta=\frac{1}{2}\lambda_{\min}(P^{1/2}G_{\dn} P^{-1/2}+ P^{-1/2}\bar G^{\top}_{\dn}P^{1/2})>0$.
	This means that $t^*\geq \frac{1-\delta^{-\mu}}{M\beta}$. Since $T(\delta)\to 0$ as $\delta\to 0$ then for
	sufficiently small $\delta$ we have $T(\delta)<t^*$, which contradicts our supposition. The proof is complete.

\subsection{The proof of Theorem \ref{thm:ch14_ext_hom_observer}}
The proof consists in the several steps: 1) we prove feasibility of algebraic systems \eqref{eq:ch14_LMI_ext_obs_Rn} and  \eqref{eq:ch14_LMI_obs_Rn}; 2) we check the homogeneity of the observer; 3) we analyze a continuity/discontinuity of the observer equation dependently on the homogeneity degree; 4) we prove finite-time stability of the error equation. On the fourth step we use  Theorem \ref{thm:hom_conv2stab}, which says that for homogeneous system with negative degree it is sufficient to guarantee a finite-time convergence of the error to zero. This convergence  is proven in three steps:
 a) the identity $\sigma(t)=\sqrt{|\xi(t)|}$ is established for $t\geq T_{\delta}$; b) it is shown that there exists $T_1>T_{\delta}$ such that either $\varepsilon_{\psi}(T_1)=\zero$ or $\xi(T_1)> \ell  \left\|\varepsilon_{\psi}(T_1)\right\|_{\bar \dn}$; c) finally, we show that the inequality $\xi(T_1)> \ell  \left\|\varepsilon_{\psi}(T_1)\right\|_{\bar \dn}$ implies that $\exists T\geq T_1$ such that $\varepsilon_{\psi}(T_1)=\zero$.

1) Due to the structure of $\{\bar A, \bar C\}$, the feasibility of the LMI \eqref{eq:ch14_LMI_ext_obs_Rn} follow from the feasibility of the LMI  \eqref{eq:ch14_LMI_obs_Rn}, which is already studied by Theorem \ref{thm:ch14_dyn_hom_obs}. Indeed,  let for $\rho>\gamma/2$ the pair  $(P,Y)$ be a solution of the LMI \eqref{eq:ch14_LMI_obs_Rn}  with  $\rho$ replaced by $2\rho$ and let $L=P^{-1}Y$.
 For any $\epsilon>0 $ there exists $\tilde p>0$ such that  $\bar P=\left(\begin{smallmatrix} \tilde p I_k & \epsilon L^{\top} P\\ \epsilon P L & \epsilon P \end{smallmatrix}\right)\succ 0$ and \vspace{-2mm}
\[
\bar P\bar G+\bar G^{\top}\bar P=\left(\begin{smallmatrix}
2\tilde p I_k & L^{\top} \epsilon PG_{\dn}\\
\epsilon G_{\dn}^{\top}	PL & \epsilon PG_{\dn}+\epsilon G_{\dn}^{\top}P
\end{smallmatrix}\right)\succ 0\vspace{-2mm}
\]
 Moreover, for
$\bar Y=\bar P\left(\begin{smallmatrix} -\tilde p I_k\\L \end{smallmatrix} \right)$ with $\tilde p>\rho$,
we have \vspace{-2mm}
\[
W:=\bar P\bar A_0\!+\!\bar A_0^{\top}\bar P+\!\bar Y\bar C\!+\!\bar C^{\top}\bar Y^{\top}\!\!+\!\rho(\bar P\bar G+\!\bar  G^{\top}\bar P)\vspace{-2mm}
\]
\[
=\left(
\begin{smallmatrix}
		- 2\tilde p (\tilde p-\rho) I_k +2\epsilon L^{\top} PL  & \tilde p C+\epsilon L^{\top} P A_{\rho}+\epsilon (1- \tilde p) L^{\top} P\\
		\tilde pC^{\top} +\epsilon A_{\rho}^{\top}P L +(\epsilon 1-\tilde p) PL & \epsilon (PA_{\rho}+A_{\rho}^{\top}P+C^{\top} L^{\top}P+ PL C)
		\end{smallmatrix}
	\right)
\]
where $A_{\rho}=A_0+\rho G_{\dn}$. Since  the pair $(P,Y)$ is selected such that $PA_{\rho}+A_{\rho}^{\top}P+C^{\top} L^{\top}P+ PL C\preceq -\rho(PG_{\dn}+G_{\dn}^{\top})\prec 0$  then for sufficiently large $\tilde p>0$ and sufficiently small $\epsilon>0$ we have $W\prec 0$.

	2) By construction, $G_{\dn}$ is anti-Hurwitz for $\nu\geq -\frac{1}{\tilde n+1}$, so
$\dn$ is a dilation in $\R^n$, $\bar \dn(s)=e^{s\bar G}$ is a dilation in $\R^{k+n}$ and $\dn_{\sigma}$ is a dilation in $\R^{k+n+1}$.
Moreover, we have \vspace{-2mm}
\[
CG_{\dn}=(1+\nu)C, \quad \bar A_0 \bar G=(\bar G+\nu I_{k+n})\bar A_0, \quad \bar C\bar G=\bar C.\vspace{-2mm}
\]	
The above identities imply that for all $\forall s\!\in\! \R$ it holds\vspace{-2mm}
\begin{equation}\label{eq:ch14_ext_obs_temp}
	C\dn(s)\!=\!e^{(1+\nu)s}C, \;\;\;\bar A_0\bar \dn(s)\!=\!e^{\nu s}\bar \dn(s)\bar A_0, \;\;\; \bar C e^{s\bar G}\!=\!e^{s}\bar C. \vspace{-1mm}
\end{equation}
Denote
$
\varepsilon_{\psi}=\left(
\begin{smallmatrix}
	\psi\\
	x-z
\end{smallmatrix}
\right)
$.

Since $	\sigma(\varepsilon_{\sigma})\!=\!\max\left\{\xi,\frac{|\bar C\varepsilon_{\psi}|^2}{2}\right\}$ then, due to the identity $\bar C\bar \dn(s)\!=\!e^{s}\bar C$, we have \vspace{-2mm}
$$
\sigma(\dn_{\sigma}(s)\varepsilon_{\sigma})=\max\left\{e^{2s}\xi,\tfrac{|\bar C\bar \dn(s)\varepsilon_{\psi}|^2}{2}\right\}=e^{2s}\sigma\vspace{-2mm}
$$
for all $s\in \R$ and for all $\varepsilon_{\sigma}\in \R^{k+n+1}$. Hence, using \eqref{eq:ch14_ext_obs_temp} we derive
$
g_{\sigma}(\dn_{\sigma}(s)\varepsilon_{\sigma})
=e^{\nu s}\dn_{\sigma}(s)g_{\sigma}(\varepsilon_{\sigma}),
$
i.e., the error equation is $\bar \dn$-homogeneous  of degree $\nu$.

3) Using properties of the matrix $G_0$ (see,  Theorem \ref{thm:ch14_dyn_hom_obs} and \cite{Polyakov_etal2016:RNC}, \cite{Zimenko_etal2020:TAC})  we conclude
$
e^{\nu s}e^{s\bar G}\to \zero  \text{ as } s\to -\infty
$
provided that $\nu>-\frac{1}{\tilde n+1}$, but
$
e^{\nu s}\bar \dn(s)\to {\rm const}  \text{ as } s\to -\infty
$
for $\nu=-\frac{1}{\tilde n+1}$. Since, by construction,  $\sigma\geq \gamma_2|\bar C\varepsilon_\psi|$  then the mapping
$
\varepsilon_{\sigma} \mapsto |\sigma|^{\nu-1}e^{\bar G \ln |\sigma|}L\bar C\varepsilon_{\psi}
$
is continuous for $\nu>-\frac{1}{\tilde n+1}$ and piece-wise continuous for $\nu=-\frac{1}{\tilde n+1}$.
The  same conclusions are obviously  valid  for the vector field $g_{\sigma}$. Therefore, the error system \eqref{eq:ch14_ext_hom_observer_error} (as well as the system \eqref{eq:ch14_ext_hom_observer}) has solutions defined at least locally in time. For $\nu=-\frac{1}{\tilde n+1}$ the solutions are understood in the sense of Filippov (see,  \cite{Filippov1988:Book}).  Moreover, for $\nu<0$ the error equation \eqref{eq:ch14_ext_hom_observer_error} is forward complete (see \cite{BhatBernstein2005:MCSS}).

4) Let us show that the error equation \eqref{eq:ch14_ext_hom_observer_error} is globally uniformly finite-time stable.
In the view of Theorem \ref{thm:hom_conv2stab},  it is sufficient to show  any trajectory of the error system reaches the origin in a finite-time and
the reaching time is a locally bounded function of the initial state.

a) For $\delta=\xi\!-\!\frac{|\bar C\varepsilon_{\psi}|^2}{2}$ we have
\begin{equation}\label{eq:ch14_temp_delta}
	\dot \delta=-\gamma |\delta|^{\frac{\nu}{2}}\delta+\left|
	\tfrac{d}{dt}\tfrac{|\psi|^2}{2}\right|- \tfrac{d}{dt}\tfrac{|\psi|^2}{2}, \quad \delta(0)=\xi(0)-\tfrac{|\bar C\varepsilon_{\psi}(0)|^2}{2}.
\end{equation}
Taking into account  $\nu<0$ we conclude that \vspace{-2mm}
\[
\xi(t)\geq \frac{|\psi(t)|^{2}}{2},\quad \forall t\geq T_{\delta},\vspace{-2mm}
\]
with $ T_{\delta}(\varepsilon_{\psi}(0),\xi(0))=\max\left\{0, -\frac{2\xi(0)- |\bar C\varepsilon_{\psi}(0)|^2}{-\nu\gamma}\right\},$
so $\sigma(t)=\sqrt{\xi(t)}$ for all $t\geq T_{\delta}(\varepsilon_{\psi}(0),\xi(0))$. Notice that (see the proof of Theorem \ref{thm:hom_conv2stab}) that there exists $M_1>0$ such that $\|\varepsilon_{\sigma}(T_{\delta})\|^{-\nu}_{\bar \dn}\leq \|\varepsilon_{\sigma}(0)\|_{\bar \dn}^{-\nu}+M_1T_{\delta}$.

b) For any $\nu\geq -\frac{1}{\tilde n+1}$ we have
\[
\Xi(r)=r^{\nu}\dn(\ln r)-r I_n\to \zero \quad \text{ as }\quad   r \to 0.
\]
Let  a constant $\ell\in (0,1)$  be selected such that
$$
\|r^{\nu}\dn(\ln r)-r I_n\|\leq \frac{\rho \beta}{2 \sqrt{2}\|\bar L\|}, \quad \forall r\in [0,\ell),
$$
where
$\beta=\frac{1}{2}\lambda_{\min}(\bar P^{1/2}\bar G\bar P^{-1/2}+\bar P^{-1/2}\bar G^{\top}\bar P^{1/2})>0$
and $\|\bar L\|=\sqrt{\lambda_{\max}(\bar L^{\top}\bar P\bar L)}$.

Let
the $\bar \dn$-homogeneous norm $\|\cdot \|_{\bar \dn}$ be induced by the weighted Euclidean norm $\|\varepsilon_{\psi}\|=\sqrt{\varepsilon_{\psi}^{\top} \bar P \varepsilon_{\psi}}$.

Let us show that for any $\varepsilon_{\sigma}(T_{\delta}) \neq \zero $
there exists $T_1\in [T_{\delta},T_{\delta}+T_{\xi}]$,
$
T_{\xi}(\varepsilon_{\sigma}(T_{\delta}))=\frac{2\left\|\varepsilon (T_{\delta})\right\|^{-\nu}_{\bar \dn}}{-\nu \rho}
$
such that  either $\varepsilon_{\psi}(T_1)=\zero$ or $\xi(T_1)> \ell  \left\|\varepsilon_{\psi}(T_1)\right\|_{\bar \dn}$.
Suppose the contrary, i.e.,  $\varepsilon_{\psi}(t)\neq \zero$ for all $t\in [T_{\delta},T_{\delta}+T_{\xi}]$ and
$
\xi(t) \leq \left\|\varepsilon_{\psi}(t)\right\|_{\bar \dn}^2,  \forall t\in [T_{\delta},T_{\delta}+T_{\xi}].
$
In this case, using the formula \eqref{eq:hom_norm_derivative_Rn} we have \vspace{-2mm}
\[
\tfrac{d}{dt} \left\| \varepsilon_{\psi}\right\|_{\bar \dn}
\!=\!\left\|\varepsilon_{\psi}\right\|_{\bar \dn}\!
\tfrac{\pi^{\top}\bar P\bar\dn\left(-\ln \left\|\varepsilon_{\psi}\right\|_{\bar \dn}\right)(	\bar A_0\varepsilon_{\psi}+\sigma^{\nu-1}\bar \dn(\ln \sigma)\bar L\bar C\varepsilon_{\psi})}{\pi^{\top}\bar P\bar G \pi}\vspace{-2mm}
\]
\[
=\!\left\| \varepsilon_{\psi}\right\|_{\bar \dn}^{1+\nu}
\tfrac{\pi^{\top}\bar P	\bar A_0 \pi+\pi^{\top}\bar P\left(\frac{\sigma}{\left\|\varepsilon_{\psi}\right\|_{\bar \dn}}\right)^{\nu-1}\bar \dn\left(\ln \frac{\sigma}{\left\| \varepsilon_{\psi}\right\|_{\bar \dn}}\right)\bar L\bar C\pi}{\pi^{\top}\bar P\bar G \pi},
\]
where
$
\pi=\bar\dn\left(-\ln \left\|\varepsilon_{\psi}\right\|_{\bar \dn}\right) \varepsilon_{\psi}
$
and the identities \eqref{eq:ch14_ext_obs_temp} are utilized on the last step.
Using  \eqref{eq:ch14_LMI_ext_obs_Rn} we derive
\[
\tfrac{d \left\| \varepsilon_{\psi}\right\|_{\bar \dn}}{dt}
= \left\|\varepsilon_{\psi}\right\|_{\bar \dn}^{1+\nu}
\left(
\tfrac{\pi^{\top}\bar P\Xi\left(\sigma/\left\| \varepsilon_{\psi}\right\|_{\bar \dn}\right)\bar L\bar C \varepsilon_{\psi}/|\sigma|}
{\pi^{\top}\bar P\bar G \pi}
-\rho\right).
\]
Since
$
\|\Xi(r)\|=\|r^{\nu}\dn(\ln r)-r I_n\|\leq \frac{\rho}{2\sqrt{2}\beta \|\bar L\|}, \forall r\in [0,\ell]
$
then
for $t\in [T_{\delta},T_{\delta}+T_{\xi}]$ we have $\sigma=\sqrt{\xi}\geq \frac{|C\varepsilon_{\psi}|}{\sqrt{2}}$, $ \frac{\sigma}{\left\|\varepsilon_{\psi}\right\|_{\bar \dn}}=\frac{\sqrt{\xi}}{\left\| \varepsilon_{\psi}\right\|_{\bar \dn}}\leq \ell$  and
$
\left\|\Xi\left(\frac{\sigma}{\left\|\varepsilon_{\psi}\right\|_{\bar \dn}}\right)\bar L\frac{\bar C \varepsilon_{\psi}}{|\sigma|}\right\|\leq \frac{\rho}{2}.
$
Therefore, taking into account $\pi^{\top}P\pi=1$ we derive
\[
\frac{d}{dt} \left\| \varepsilon_{\psi}\right\|_{\bar \dn}\leq -\frac{\rho}{2} \left\| \varepsilon_{\psi}\right\|_{\bar \dn}^{1+\nu}, \forall t\in [T_{\delta},T_{\delta}+T_{\xi}],
\]
but the latter implies there exists $T_1\in [T_{\delta},T_{\delta}+T_{\xi}]$ such that $\varepsilon_{\psi}(T_1)=\zero$ and
we obtain the contradiction. Therefore, either $\varepsilon_{\psi}(T_1)=\zero$
or $\xi(T_1)>\ell \|\varepsilon_{\psi}(T_1)\|_{\bar \dn}^2$. In the first case, the state $\varepsilon_{\psi}(t)$ will remain zero at least till the time instant $T_1+\frac{\xi^{-\nu}(T_1)}{-\nu\gamma}$ then $\xi(t)$ will reach zero too, i.e.,
the error variable $\varepsilon_{\sigma}(t)$ reaches zero at a finite  instant of time being a locally bounded function of $\varepsilon_{\sigma}(0)$. Notice that (see the proof of Theorem \ref{thm:hom_conv2stab}) there exists
$M_2>0$ such that $\xi^{-\nu/2}(T_1)\leq \|\varepsilon_{\sigma}(T_1)\|_{\dn_{\sigma}}:=\xi^{-\nu/2}(T_1)+\|\varepsilon_{\psi}(t)\|_{\bar \dn}\leq  \|\varepsilon_{\sigma}(0)\|_{\bar \dn}+M_2 T_1. $

c) Let us study the case $\xi(T_1)>\ell \|\varepsilon_{\psi}(T_1)\|_{\bar \dn}^2$ and consider
$
\epsilon= e^{-\frac{1}{2}\bar G\ln\xi}\varepsilon_{\psi},
$
which is well defined for $t\ge T_1$. Notice,
\[
\| \dn(\ln \ell)  \epsilon(T_1)\|_{\bar \dn}<1 \quad \Leftrightarrow \quad
\xi(T_1)>\ell \|\varepsilon_{\psi}(T_1)\|_{\bar \dn}^2.
\]
and (due to the estimates \eqref{eq:rel_norm_and_hom_norm_Rn}, \eqref{eq:dilation_rates_in_Rn}) we have
\[
\ell^{\alpha}\| \epsilon(T_1)\|\leq \|\dn(\ln \ell) \epsilon(T_1)\|<1,
\]
where $\alpha=\frac{1}{2}\lambda_{\max}(\bar P^{1/2}\bar G\bar P^{-1/2}+\bar P^{-1/2}\bar G^{\top}\bar P^{1/2})>0$.
Using the identities
$
\bar C\varepsilon_{\psi}=
\xi^{\frac{1}{2}}\bar C\epsilon
$ and
$
\bar C\bar A_0\varepsilon_{\psi}=
\xi^{\frac{1+\nu}{2}}	\bar C\bar A_0\epsilon
$
we derive \vspace{-2mm}
\[
\varepsilon_{\psi}^{\top}\bar C^{\top}\!\left(\bar C (\bar A_0+|\sigma|^{\nu}\tilde L\bar C)\varepsilon_{\psi}\right)=\xi^{1+\frac{\nu}{2}}
\epsilon^{\top}\bar C^{\top}(\bar C\bar A_0+
L\bar C)\epsilon.\vspace{-2mm}
\]
Since $\xi\geq \frac{|\bar C\varepsilon_{\psi}|^{2}}{2}\Rightarrow 1\geq \frac{|\bar C\epsilon|^{2}}{2}$ then\vspace{-2mm}
\[
\dot \xi=(h(\epsilon)-\gamma)\xi^{1+\frac{\nu}{2}},\quad \forall t\geq T_1\geq T_{\delta},\vspace{-2mm}
\]
\[
h(\epsilon)\!=\!|\epsilon^{\top}\!\bar C^{\top}\!\bar C(\bar A_0+
\bar L\bar C)\epsilon|+\gamma-\gamma\left(1-\tfrac{|\bar C\epsilon|^{2}}{2
}\right)^{1+\frac{\nu}{2}}\geq 0.
\]
For $1\geq \frac{|\bar C\epsilon|^{2}}{2}$  and $-2<\nu<0$ we have
$
1 -\left(1-\frac{|\bar C\epsilon|^{2}}{2}\right)^{1+\frac{\nu}{2}}\leq  \frac{|\bar C\epsilon|^{2}}{2}.
$
Therefore,  $\exists \kappa>0$ (dependent only of $\bar P, \bar C, \bar L, \bar A_0$ and $\gamma>0$) such that
$
1\geq \frac{|\bar C\epsilon|^{2}}{2} \Rightarrow h(\epsilon) \leq \kappa \|\epsilon\|^2
$
and
\begin{equation}\label{eq:ch14_ext_obs_temp_xi}
	\dot \xi\leq \xi^{1+\frac{\nu}{2}}(q^{-1}\kappa \|\epsilon\|^2-\gamma), \quad \forall t\in [T_1,T).
\end{equation}
The time derivative of $\epsilon$ satisfies the equation:\vspace{-2mm}
\[
\dot \epsilon\!=-\tfrac{1}{2}\tfrac{\dot \xi}{\xi}
\bar G\epsilon+e^{-\frac{1}{2}\bar G\ln \xi}(\bar A_0\varepsilon_{\psi}+\xi^{\frac{\nu-1}{2}}e^{\frac{1}{2}\bar G \ln \xi}L\bar C\varepsilon_{\psi})\vspace{-2mm}
\]
\[
=\xi^{\frac{\nu}{2}} \left(\bar A_0+L\bar C+(0.5\gamma-
h(\epsilon)) \bar G\right)\epsilon.
\]
Considering $V=q\epsilon^{\top}\bar P \epsilon$  with  $q=\frac{\kappa(1-0.5\nu)}{\beta(\rho-0.5\gamma)}$  we derive\vspace{-2mm}
\[
\dot V=
2q\xi^{\frac{\nu}{2}} \epsilon^{\top}\bar P\left(\bar A_0+L\bar C+
\tfrac{\gamma-h(\epsilon)}{2}  \bar G\right)\epsilon\vspace{-3mm}
\]
\[
\leq
-\left(0.5h(\epsilon)+\rho -0.5\gamma\right)\xi^{\frac{\nu}{2}}
2q\epsilon^{\top}\bar P\bar G \epsilon,
\]
where the inequality \eqref{eq:ch14_LMI_ext_obs_Rn} is utilized on the last step.
Since  $h(\epsilon)\geq 0$  then for   $\rho>\frac{\gamma}{2}$ we have \vspace{-2mm}
\begin{equation}\label{eq:ch14_V_dot_temp}
	\dot V
	\leq  -\beta (\rho-0.5\gamma)\xi^{\frac{\nu}{2}} V.\vspace{-2mm}
\end{equation}
The latter means that the function $t\mapsto V(t)$ is strictly decreasing on $[T_1,T)$ and \vspace{-2mm}
$$
\|\epsilon(t)\|\leq\|\epsilon(T_2)\|,  \quad \forall T_2\in(T_1,T), \quad \forall t\in[T_2,T),\vspace{-2mm}
$$
where $\|\epsilon\|=\sqrt{\epsilon^{\top}\bar P\epsilon}$ is the weighted Euclidean norm  in $\R^{k+n}$.
Considering
$
W=\xi^{1-\frac{\nu}{2}}e^{V}
$
we derive \vspace{-2mm}
\[
\begin{array}{c}
	\dot W=(1-0.5\nu)\xi^{-\frac{\nu}{2}}\dot \xi e^{V}+
\xi^{1-\frac{\nu}{2}} e^{V}\dot V
\\
\leq   (1-0.5\nu)e^{V}\xi(q^{-1}\kappa V-\gamma)-\beta (\rho-0.5\gamma) \xi e^{V}
\\
=-
(1-0.5\nu)\gamma \xi e^{V}=
-(1-0.5\nu)\gamma\xi^{0.5\nu}W.
\end{array}\vspace{-2mm}
\]
Therefore, we have the following system of inequalities  \vspace{-2mm}
\[
\left\{
\begin{array}{l}
	\dot V\leq -\beta(\rho-0.5\gamma) W^{\frac{0.5\nu}{1-0.5\nu}} e^{\frac{-0.5\nu}{1-0.5\nu}V}V,\\
	\dot W\leq -(1-0.5\nu)\gamma e^{\frac{-0.5\nu}{1-0.5\nu} V} W^{1+\frac{0.5\nu}{1-0.5\nu}},
\end{array}
\right. \quad \forall t\in[T_1,T)\vspace{-2mm}
\]
Notice for $\nu=-1/(\tilde n+1)$ the above inequalities hold almost everywhere in the view of Filippov solutions \cite{Filippov1988:Book}.
Taking into account $V\geq 0 \Rightarrow e^{V}\geq 1$ we conclude that\vspace{-2mm}
\[
\exists T\in \left(T_1,\tfrac{(1-0.5\nu)W^{-\frac{0.5\nu}{1-0.5\nu}}(T_1)}{(1-0.5\nu)\gamma (-0.5\nu)}\right),\vspace{-2mm}
\]
such that
$W(t)\to0$ and $V(t)\to 0$ as $t\to T$
or, equivalently, $\epsilon(t)\to \zero$ and $\xi(t)\to0$ as $t\to T$.
Taking into account $\|\varepsilon_{\psi}(t)\|_{\bar \dn}=\xi(t)\|\epsilon(t)\|_{\bar\dn}$
and
\vspace{-2mm} $$
\tfrac{(1-0.5\nu)W^{-\frac{0.5\nu}{1-0.5\nu}}(T_1)}{(1-0.5\nu)\gamma (-0.5\nu)}\leq
\tfrac{(1-0.5\nu)\left(\xi(T_1)^{1-0.5\nu} e^{\ell^{-2\alpha}}\right)^{-\frac{0.5\nu}{1-0.5\nu}}}{(1-0.5\nu)\gamma (-0.5\nu)},\vspace{-2mm}
$$
we conclude that any trajectory of the error equation reaches the origin in a finite time and this time is a locally bounded function of the initial state. Using Theorem \ref{thm:hom_conv2stab} we complete the proof.

\subsection{The proof of Corollary \ref{cor:HOSM}}
Let us consider the  extension  of the error equation:\vspace{-2mm}
\begin{equation}
	\dot \varepsilon_{\sigma}\!\in\!G(\epsilon_{\sigma})=g(\varepsilon_{\sigma})+\left(\begin{smallmatrix} \zero\\ E\\
		0\end{smallmatrix}\right)\mathcal{B}_{\bar q },\vspace{-2mm}
\end{equation}
 where $\mathcal{B}_{\bar q }=\{x\in \R^n: |x|\leq \bar q \}$ is the Euclidean ball in $\R^n$ of the radius $\bar q>0$. Since
 $
 (\tilde n+1)G_{\dn}E\!=\!E  \Rightarrow G_{\dn}E\!=\!-\nu E \Rightarrow \dn(s)E\!=\!e^{-\nu s}E, \forall s\!\in\! \R
 $
 then the set-valued vector field $G$ is
 $\dn$-homogeneous:
 $G(\dn_{\sigma}(s)x)\!=\!g(\dn_{\sigma}(s) \epsilon_{\sigma})\!+\!\left(\!\begin{smallmatrix} \zero\\ E\\
 	0\end{smallmatrix}\!\right)\!\mathcal{B}_{\bar q }
 $
 $\!=\!e^{\nu s}\dn_{\sigma}(s) \left(g(\epsilon_{\sigma})\!+\!\left(\!\begin{smallmatrix} \zero\\ E\\
 	0\end{smallmatrix}\right)\! \mathcal{B}_{\bar q }\!\right )$
  $=e^{\nu s} \dn_{\sigma}(s)G(\epsilon_{\sigma}).$
 Since, for $\bar q=0$ the error system  is finite-time stable then,  by Zubov-Rosier Theorem \cite{Zubov1958:IVM}, \cite{Rosier1992:SCL},  there exists a $\dn_{\sigma}$-homogeneous Lyapunov function $V\in C(\R^{k+n+1})\cap C^{\infty}(\R^{k+n+1}\backslash\{\zero\})$ of degree $-\nu$ such that
 $
 \dot V(\epsilon_{\sigma})=-\kappa, \quad \forall \epsilon_{\sigma}\neq 0,
 $
 for some $\kappa >0$.
In this case, for the perturbed case $\bar q\neq 0$ we have\vspace{-2mm}
 \[
  \dot V(\epsilon_{\sigma})\le-\kappa+\tfrac{\partial V(\epsilon_{\sigma})}{\partial \epsilon_{\sigma}} \left(\!\begin{smallmatrix} \zero\\ E\\
  	0\end{smallmatrix}\!\right) q\vspace{-2mm}
 \]
 with $q\in \mathcal{B}_{\bar q }$. Since
 the partial derivative of the homogeneous function is the $\dn_{\sigma}$-homogeneous \cite{Polyakov2020:Book}
 $
 \left.\frac{\partial V(\zeta)}{\partial \epsilon_{\zeta}}\right|_{\zeta=\dn_{\sigma}(s)\epsilon_{\sigma}}\dn_{\sigma}(s)= e^{-\nu s}\frac{\partial V(\epsilon_{\sigma})}{\partial \epsilon_{\sigma}}, \forall s\in \R
 $ then taking $s=-\ln \|\epsilon_{\sigma}\|_{\dn_{\sigma}}$ we derive \vspace{-2mm}
 $$\left | \tfrac{\partial V(\epsilon_{\sigma})}{\partial \epsilon_{\sigma}} \left(\!\begin{smallmatrix} \zero\\ E\\
 	0\end{smallmatrix}\!\right)\right|\leq \bar V'=\sup_{\|\zeta\|=1} \left | \tfrac{\partial V(\zeta)}{\partial \zeta} \left(\!\begin{smallmatrix} \zero\\ E\\
 	0\end{smallmatrix}\!\right)\right|<+\infty\vspace{-2mm}$$
 where the identity  $e^{\nu s}\dn_{\sigma}(s)E=E,\forall s\in \R$ is utilized on the last step. Therefore, the error system remains finite-time stable for $\bar q\leq \frac{\kappa}{\bar V'}$.

\bibliographystyle{plain}

\end{document}